\documentclass[11pt]{article}
\usepackage{epsfig} 
\usepackage{amssymb}
\newcommand{\sect}[1]{ \section{#1} \setcounter{equation}{0} }

\newcommand{\partialslash}{\partial \! \! \! /}
\newcommand{\half}{\mbox{\small{$\frac{1}{2}$}}}

\newcommand{\MSbar}{\overline{\mbox{MS}}}

\newcommand{\Nf}{N_{\!f}}

\begin{document}
\title{Higher dimensional higher derivative $\phi^4$ theory}
\author{J.A. Gracey \& R.M. Simms, \\ Theoretical Physics Division, \\ 
Department of Mathematical Sciences, \\ University of Liverpool, \\ P.O. Box 
147, \\ Liverpool, \\ L69 3BX, \\ United Kingdom.} 
\date{}
\maketitle 

\vspace{5cm} 
\noindent 
{\bf Abstract.} We construct several towers of scalar quantum field theories
with an $O(N)$ symmetry which have higher derivative kinetic terms. The
Lagrangians in each tower are connected by lying in the same universality class
at the $d$-dimensional Wilson-Fisher fixed point. Moreover the universal theory 
is studied using the large $N$ expansion and we determine $d$-dimensional 
critical exponents to $O(1/N^2)$. We show that these new universality classes
emerge naturally as solutions to the linear relation of the dimensions of the
fields deduced from the underlying force-matter interaction of the universal
critical theory. To substantiate the equivalence of the Lagrangians in each 
tower we renormalize each to several loop orders and show that the 
renormalization group functions are consistent with the large $N$ critical 
exponents. While we focus on the first two new towers of theories and 
renormalize the respective Lagrangians to $16$ and $18$ dimensions there are an
infinite number of such towers. We also briefly discuss the conformal windows 
and the extension of the ideas to theories with spin-$\half$ and spin-$1$
fields as well as the idea of lower dimension completeness.  

\vspace{-20cm}
\hspace{13cm}
{\bf LTH 1134}

\newpage 

\sect{Introduction.}

The Wilson-Fisher fixed point of scalar quantum field theories has provided a
remarkable basis for tackling a variety of different problems in physics,
\cite{1,2,3,4}. The most widely known example is the use of the Wilson-Fisher 
fixed point underlying the Ising model as well as the superfluid phase 
transition, dilute polymer solutions and the Heisenberg ferromagnet. 
Information about the properties of their phase transitions can be accessed by 
the continuum scalar quantum field theory with a quartic interaction. When 
endowed with an $O(N)$ symmetry the $N$~$=$~$1$ case corresponds to the Ising 
model whereas the ferromagnet is described by the value of $N$~$=$~$3$. Equally
dilute polymer solutions and superfluidity correspond to the cases of
$N$~$=$~$0$ and $2$ respectively. The connection with the Heisenberg magnet 
is perhaps remarkable, for instance, in that information on the phase 
transition in three dimensions can be obtained by renormalizing the $O(N)$ 
$\phi^4$ scalar theory in {\em four} dimensions. Central to this connection is 
the Wilson-Fisher fixed point and the underlying critical renormalization group
equation, \cite{1,2,3,4}, which is a core property in $d$-dimensions. Moreover
underlying the transition is a universal quantum field theory which is central
to the theories at the Wilson-Fisher fixed point. Put another way in the 
approach to three dimensions through the $\epsilon$ expansion, where the 
spacetime dimension is $d$~$=$~$4$~$-$~$2\epsilon$, only the quartic operator 
present in the $O(N)$ scalar theory is relevant. In practical terms to obtain 
accurate information on the phase transition properties, one would have to know
the renormalization group functions of $O(N)$ $\phi^4$ theory to a large loop 
order in four dimensions where it is renormalizable. This has been achieved 
over many years in the advance to five loop renormalization group functions in 
the modified minimal subtraction ($\MSbar$) scheme in \cite{5,6,7,8,9,10}. More
recently these calculations have been extended to {\em six} loops in 
\cite{11,12,13,14}. From a bigger perspective our understanding of the 
universality property of this particular Wilson-Fisher fixed point has been 
extended above and beyond four dimensions in recent years. 

Similarly the connection to the $O(N)$ nonlinear sigma model, which is 
renormalizable in two dimensions, is already well-established and the extension
to six dimensions was achieved in \cite{15,16,17} based on the early work of 
\cite{18,19} and eight dimensions in \cite{20}. The connections were 
constructed by explicit perturbative renormalization of the respective higher 
dimensional $O(N)$ symmetric scalar theories and, in addition, knowledge of the
critical exponents of the universal theory in arbitrary spacetime dimensions. 
The latter is possible through an expansion akin to conventional perturbation 
theory but where the expansion parameter is dimensionless similar to the 
perturbative coupling constant in the critical dimension of a theory. 
Specifically this is the parameter $1/N$ when an $O(N)$ symmetry is present. 
When $N$ is large then $1/N$ is small and this can be regarded as a 
perturbative parameter which unlike a coupling constant is dimensionless in any
spacetime dimension. What is also remarkable is that the core critical 
exponents have been determined at a fixed point, such as the Wilson-Fisher one,
to {\em three} orders in $1/N$, \cite{21,22,23,24}. Also these exponents, which
are renormalization group invariants and known as exact functions of $d$ at 
each order in $1/N$, have a correspondence with the renormalization group 
functions of the critical spacetime dimension of each of the theories in the 
same universality class. Therefore evaluating the exponents in the $\epsilon$ 
expansion near each critical dimension provides information on each set of 
renormalization group functions of the theory in that critical dimension. 
Equally knowledge of the latter at a particular loop order establishes the 
connection of the six and higher dimensional quantum field theories with the 
same universality class as $O(N)$ $\phi^4$ theory and the $O(N)$ nonlinear 
sigma model. From a Lagrangian point of view what is central in the universal 
theory is a core interaction which in essence has a force-matter structure of 
the form $\sigma \phi^i \phi^i$. Here $\phi^i$ is the $O(N)$ scalar field and 
$\sigma$ is a second scalar field representing the force mediating particle. 
This interaction drives the Wilson-Fisher fixed point dynamics and is the only 
interaction in the universal theory used for the computation of the 
$d$-dependent critical exponents in the $1/N$ expansion. Although each 
Lagrangian in the critical spacetime dimension will have this same core 
interaction they differ in the other interactions. By performing a dimensional 
analysis of the Lagrangians of the higher dimensional theories one can abstract
the essence of those extra interactions. They will only involve $\sigma$ 
self-interactions which will differ from dimension to dimension. We will term 
these as spectator interactions in contrast to core, due to the sense that they
only participate or are activated in one specific dimension. Their absence in 
the Lagrangian at its critical dimension would render the theory 
non-renormalizable and hence not connected to the universal theory. This 
resulting tower of theories in the $O(N)$ scalar theory universality tree is 
now well-established from high order perturbative and $1/N$ computations. For 
example, see \cite{15,16,17,20}.

One of the more recent applications of the $1/N$ formalism of \cite{21,22,23}
is its connection with $d$-dimensional conformal field theories. Earlier work
in this direction was provided in \cite{25,26,27,28} for example. In this 
approach it is hoped that the ideas of two dimensional conformal field theories
can be developed in dimensions greater than two and the $1/N$ tool assists this
activity. Equally a less well explored aspect of scalar and other conformal
field theories is that where the kinetic term of the scalar field involves
higher derivatives. One recent study of tying these two threads together was
in \cite{29,30,31}. Although \cite{29,30,31} focussed on the free $\Box^2$ 
conformal field theories and its symmetry algebra, this has to be appreciated 
first before the extension to interacting theories. Moreover that work was 
motivated by connections with AdS/CFT ideas. It is therefore the purpose of 
the article to contribute to this debate by examining scalar field theories 
using the $1/N$ expansion formalism of \cite{21,22} in a new way. We will show 
how one can construct new towers of interacting theories based around distinct 
universal theories and connected by a Wilson-Fisher fixed point. As the initial
distinction is at the level of the $\phi^i$ field kinetic term there will in
principle be a countably infinite number of new universality classes but we 
will concentrate on the first few of these in detail. As the starting point of 
our description of the Wilson-Fisher fixed point is $O(N)$ $\phi^4$ scalar 
theory we will term these new towers of theories as higher derivative $\phi^4$ 
theories. Part of our construction will involve computing several large $N$ 
critical exponents for the first two new towers explicitly using the methods of 
\cite{21,22}. Then to establish the connection with the respective field 
theories in specific dimensions we will construct the Lagrangians and 
renormalize them to several loop orders. In some cases this will be to three 
loops and the tower will extend to $16$ and $18$ spacetime dimensions 
respectively. 

Having these explicit connections with higher derivative theories can serve at 
the very least as an alternative forum or laboratory to test new ideas in 
connections with conformal field theories as there is an underlying 
Wilson-Fisher fixed point. Our study will not be the first in this area. For 
instance, higher derivative nonlinear sigma models were considered in 
\cite{32,33} where various one loop $\beta$-functions were computed. More 
recently, similar ideas have been put forward in \cite{34,35}. Although these 
provide valuable insights, our analysis has the benefit of cross-connecting 
theories dimensionally. Our main focus thoughout will be on scalar theories but
we will not restrict the discussion solely to these. Theories which have 
similar $1/N$ expansions in the formalism of \cite{21,22} will be discussed in 
part. The key two are the $O(N)$ Gross-Neveu model, \cite{36}, and the 
non-abelian Thirring model, \cite{37,38}. These two theories which are 
renormalizable in two dimensions, serve as the base Lagrangian in the large $N$
formalism and are in the same universality class as four dimensional 
Gross-Neveu-Yukawa, \cite{39}, and Quantum Chromodynamics (QCD) respectively. 
In each case the matter field is fermionic but this is no obstacle to having a 
parallel higher derivative kinetic term which we will discuss here. Indeed an 
early construction of a renormalizable extension of a fermion field involving
higher derivatives was noted in \cite{40}. A final issue which we will touch on
as well, since it will be a manifestly new feature of the higher derivative 
$\phi^4$ theories, is that of what we will call lower dimension completeness. 
The extension of $\phi^4$ theory to six and higher dimensions at the standard 
Wilson-Fisher fixed point is called ultraviolet completion. As the higher 
derivative $O(N)$ $\phi^4$ theories will have a critical dimension greater than
four a natural question to ask is what would be the lower dimensional theory in
the higher derivative universality class. In some sense this in the opposite 
direction to the ultraviolet completion in terms of dimensions.

The article is organized as follows. We review the essential aspects of the
large $N$ method of \cite{21,22,23} as well as introducing the new solutions
which lead to the higher derivative $O(N)$ theories in the following section.
Once these new universal large $N$ Lagrangians have been constructed we compute
various $d$-dimensional critical exponents in the $1/N$ expansion in section
$3$. In order to compare the information contained within these exponents
section $4$ is devoted to constructing the towers of ultraviolet complete
Lagrangians for the two main extensions of higher derivative Lagrangians we
concentrate on. The explicit renormalization of these Lagrangians is carried 
out in the subsequent section. We also briefly discuss the status of the
conformal windows in these theories there. Section $6$ lays the groundwork for 
the extension of these ideas to theories with fermions which are also
accessible via the large $N$ expansion. As the higher derivative $\phi^4$ core
Lagrangians are renormalizable in a dimension higher than the canonical one in
four dimensions there is the possibility of having lower dimensional theories
in the same universality class. These Lagrangians are tentatively explored in 
section $7$ while we provide conclusions in section $8$.

\sect{Background.}

In order to motivate the connection which a higher derivative $\phi^4$ theory
has with its canonical four dimensional counterpart we review the basic 
Lagrangian within the large $N$ formalism. The four dimensional renormalizable
Lagrangian is 
\begin{equation}
L^{(4)} ~=~ \frac{1}{2} \partial_\mu \phi^i \partial^\mu \phi^i ~+~
\frac{1}{8} g_1^2 \left( \phi^i \phi^i \right)^2 
\label{lag4}
\end{equation}
where $g_1$ is a dimensionless coupling in four dimensions and throughout
$1$~$\leq$~$i$~$\leq$~$N$. While this version is the one widely used to
construct the renormalization group functions, \cite{5,6,7,8,9,10}, the 
interaction can be rewritten in terms of an auxiliary field $\sigma$ to produce
the equivalent Lagrangian 
\begin{equation}
L^{(4)} ~=~ \frac{1}{2} \partial_\mu \phi^i \partial^\mu \phi^i ~+~
\frac{g_1}{2} \sigma \phi^i \phi^i ~-~ \frac{1}{2} \sigma^2 ~. 
\label{lag44}
\end{equation}
It is this version of the original scalar quartic theory which is the starting 
point for the large $N$ construction provided by Vasil'ev et al,
\cite{21,22,23}. In this approach the critical exponents of the theory are
determined directly from the field theory as a series in powers of $1/N$
without using the perturbative renormalization group functions. Defining the
full dimensions of the fields $\phi^i$ and $\sigma$ by $\alpha$ and $\beta$
respectively then, \cite{21,22},
\begin{equation}
\alpha ~=~ \mu ~-~ 1 ~+~ \half \eta ~~~,~~~ \beta ~=~ 2 ~-~ \eta ~-~ \chi
\label{albet}
\end{equation}
where $d$~$=$~$2\mu$. The canonical dimensions of the fields are determined by
a dimensional analysis of the Lagrangian with the proviso that the action $S$
is dimensionless. The spacetime dimension enters via the $d$-dimensional 
measure associated with the relation between the Lagrangian and $S$. The other
quantities, $\eta$ and $\chi$, correspond to critical exponents and represent 
or are a measure of the quantum corrections. They are referred to as anomalous
dimensions. The exponent associated with $\phi^i$ is $\eta$ whereas $\chi$ is 
the anomalous dimension of the vertex operator $\sigma \phi^i \phi^i$. 

Strictly in the Vasil'ev et al definition of the canonical dimensions the 
Lagrangian 
\begin{equation}
\bar{L}^{(4)} ~=~ \frac{1}{2} \partial_\mu \phi^i \partial^\mu \phi^i ~+~
\frac{1}{2} \bar{\sigma} \phi^i \phi^i ~-~ \frac{1}{2g_1^2} \bar{\sigma}^2 
\label{lag44n}
\end{equation}
is used, \cite{21,22}, where the $\sigma$ field is rescaled by a power of the 
coupling. In this version of $L^{(4)}$ the $\bar{\sigma} \phi^i \phi^i$ 
interaction is central and drives the universality class across the dimensions 
where the $O(N)$ nonlinear sigma model, $O(N)$ $\phi^4$ theory and $O(N)$ 
$\phi^3$ theory in six dimensions, \cite{15,16}, are all critically equivalent 
at the Wilson-Fisher fixed point together with higher dimensional extensions, 
\cite{20}. Therefore $L^{(4)}$ is the version which was used in \cite{21,22,23}
to define the critical point propagators of the two fields which are
\begin{equation}
\langle \phi^i(x) \phi^j(y) \rangle ~\sim~ 
\frac{A \delta^{ij}}{((x-y)^2)^\alpha} ~~~,~~~ 
\langle \sigma(x) \sigma(y) \rangle ~\sim~ \frac{B}{((x-y)^2)^\beta} 
\label{critprop}
\end{equation}
in coordinate space where $A$ and $B$ are $x$ and $y$ independent amplitudes.
Within computations they always appear in the combination $z$~$=$~$A^2B$. It is
important to recognise that these are the propagators in the asymptotic
approach to the critical point which is why they do not involve equality 
symbols. As noted in \cite{22} there are corrections to scaling. While the 
canonical exponents are determined by dimensional arguments expressions for 
$\eta$ and $\chi$ are deduced by solving the skeleton Dyson-Schwinger equations
at criticality where the integrals are regularized analytically, 
\cite{21,22,23}. This is a crucial point as one is in effect performing a 
perturbative expansion in the anomalous dimension of the vertex operator. 
Therefore the regularization is formally introduced by the shift
$\chi$~$\to$~$\chi$~$+$~$\Delta$ where $\Delta$ is the regularizing parameter,
\cite{21,22,23}. The advantage of this is that the usual dimensional 
regularization of perturbation theory is not used. So the underlying Feynman 
integrals are evaluated as {\em exact} functions of $d$ at each order in the 
$1/N$ expansion. The notation introduced in \cite{21} reflects this since
\begin{equation}
\eta ~=~ \sum_{i=1}^\infty \frac{\eta_i(\mu)}{N^i} 
\end{equation}
with similar expansions for other quantities such as $\chi$ and $z$. In other 
words the coefficients of the expansion parameter are functions of the
spacetime dimension. 

At this point it is worth making the connection of the large $N$ formulation
with the conventional perturbative approach. The exponents, such as $\eta$
which has been determined to $O(1/N^3)$ in \cite{21,22,23}, are renormalization
group invariants and thus are more fundamental than the renormalization group 
functions themselves. However both are related since the evaluation of a 
renormalization group function at a value of the coupling defined by a zero of 
the $\beta$-function defines a critical exponent. In other words we have 
relations such as
\begin{equation}
\eta ~=~ \gamma_\phi(g_c)
\label{etadef}
\end{equation}
where $\gamma_\phi(g)$ is the renormalization group anomalous dimension of the 
field $\phi^i$. Here $g_c$ is the critical coupling defined by 
$\beta(g_c)$~$=$~$0$ and we use $g$ to represent either a single coupling, such
as $g_1$ in (\ref{lag4}), or a vector of couplings for theories where there is 
more than one coupling constant which will arise later. While this outlines the
essence of the critical point renormalization group relevant to the large $N$ 
the final aspect which makes the connection is that the solution to 
$\beta(g_c)$~$=$~$0$ is determined from the $d$-dimensional $\beta$-function. 
One of the solutions to this equation will be the Wilson-Fisher fixed point. In
carrying out the classical dimensional analysis of the Lagrangian above to 
determine the canonical dimensions of the field we omitted the analysis of the 
coupling constant. Ordinarily it derives from the interaction in (\ref{lag4}) 
but in the large $N$ context one uses the quadratic term in (\ref{lag44n}). In 
either situation one finds that the dimension of the coupling is $(d-4)$ in 
$d$-dimensions. In other words the critical dimension, from the point of view 
of renormalizability, is four and in that dimension the coupling is 
dimensionless. This is reflected in the $d$-dimensional $\beta$-function whose 
first term will be proportional to $(d-4)g$ with the actual constant depending 
on the convention used to define the $\beta$-function. The dimensional 
dependence of the remaining part of the $\beta$-function depends on the 
renormalization scheme which is chosen. In general the coefficients of the one 
and higher loop terms in the series will be of the form $A_n$~$+$~$B_n (d-4)$ 
where $A_n$ is related to the residue of the simple pole in the regularizing 
parameter in the coupling constant renormalization constant while $B_n$ is
connected to its finite part. In the strictly four dimensional $\beta$-function
only the sequence of numbers $A_n$ would be present and the first term of 
$\beta(g)$ would involve $A_1$. As an aside we note that this description of 
the structure of the $d$-dimensional $\beta$-function applies to all the other 
renormalization group functions. The finite part of the renormalization 
constant appears in the same way in the $d$-dimensional expression. The 
relevance of these finite parts appearing in the renormalization group 
functions emerges when one determines the underlying critical exponents in 
whatever scheme one has performed the perturbative renormalization. Solving 
$\beta(g_c)$~$=$~$0$ in $d$-dimensions and determining $\gamma_\phi(g_c)$ the 
overall expression has to be scheme independent as the critical exponents are 
renormalization group invariants. Omitting the respective $B_n$ sequences from 
the renormalization group functions would lead to a contradiction of this. For 
explicit three loop examples we refer the interested reader to \cite{41} where 
the $\beta$-functions of QCD were examined in this context in the $\MSbar$ and 
the momentum subtraction (MOM) schemes of Celmaster and Gonsalves, 
\cite{42,43}.

We have discussed the background to the $d$-dimensional renormalization group
functions at length as it is relevant to the relation to the $d$-dimensional
large $N$ critical exponents of \cite{21,22,23}. In our discussion we have been
careful in our description. First, the role of the renormalization scheme is 
relevant in the $d$-dimensional renormalization group functions. However we 
will not comment any further on this here as throughout we will use the
$\MSbar$ scheme for all our perturbative computations. We recall that in the 
$\MSbar$ scheme the renormalization constants are defined at a specific 
subtraction point by removing only those terms which are divergent with respect
to the parameter of the regularization. This naturally leads to the second 
aspect of our discussion which is that at no point did we refer to a specific 
regularization let alone dimensional regularization which is the one widely 
used for explicit perturbative renormalization at higher loop order. In this 
regularization the critical dimension of the field theory is replaced by a 
complex variable $d$ which is then written in our four dimensional case as 
$d$~$=$~$4$~$-$~$2\epsilon$ where $\epsilon$ is the regularizing parameter of
dimensional regularization. The use of $d$ here differs in sense from that used
in our earlier description of the $d$-dimensional renormalization group 
functions. For instance, if one used a different regularization to dimensional 
regularization, such as lattice regularization, and renormalized in a MOM 
scheme, for example, then the finite parts of the renormalization constant 
represented by $B_n$ would appear in the corresponding $d$-dimensional 
renormalization group functions. In other words there is a clear demarcation 
between use of dimensional regularization and the construction of the 
renormalization group functions in $d$-dimensional spacetime. In the context of
the universal theory represented by the Lagrangian used for the large $N$ 
expansion, (\ref{lag44n}), it is the $d$-dimensional renormalization group 
functions which are used to compare with the exponents defined by the relations
like (\ref{etadef}). We have discussed this issue at length due to its subtle 
nature. Therefore, we summarize the situation by clarifying that the exponents 
derived in the large $N$ expansion for the universal theory are functions of 
the actual spacetime dimension $d$ itself, which is completely arbitrary, and
the exponents are derived in an {\em analytically} regularized Lagrangian. By 
contrast in perturbation theory using $\epsilon$ within dimensional 
regularization it is regarded as being a small quantity.

The connection between the large $N$ exponents and the perturbative 
renormalization group functions comes through the definitions such as
(\ref{etadef}). The quantities in (\ref{etadef}) are functions of $d$ and $N$
and can be expanded as a double Taylor series in the respective parameters
$\epsilon$ and $1/N$ which are small and dimensionless. Here $\epsilon$ is not 
being used as a regularizing parameter in the strict sense as it would be in 
dimensional regularization as the exponents are the outcome of the large $N$
formalism with the universal theory analytically regularized for all $d$. 
Instead we have set $d$~$=$~$D$~$-$~$2\epsilon$ where $D$ is an integer and 
represents the critical dimension of one of the theories in the tower connected
by the universal interaction. If one computes the renormalization group 
functions for a specific theory with critical dimension $D$ in that tower then 
expands the exponent in powers of $\epsilon$ at each order in $1/N$ for both 
the large $N$ derived approximation to the exponent and the perturbative 
renormalization group function determined at the critical $D$ Wilson-Fisher 
fixed point one finds total agreement of both expansions. Of course in 
practical terms one only knows several orders in the respective expansions but 
this is usually sufficient to establish consistency. As such this approach has 
become a standard check on explicit high loop order perturbative expansions 
where the theory has an $O(N)$ or similar symmetry with a {\em dimensionless} 
parameter. Equally it has been used to determine coefficients in the
polynomial of $N$ at higher loop orders which have not been evaluated
perturbatively. Recent examples are the determination of the six loop $O(N)$ 
$\phi^4$ $\beta$-function and other renormalization group functions in the 
$\MSbar$ scheme, \cite{5,6,7,8,9,10,11,12,13,14}. The $O(\epsilon^6)$ terms of 
the derived exponents of \cite{11,12,13,14} were in exact agreement with the 
various $O(1/N^2)$ and $O(1/N^3)$ exponents of \cite{21,22,23,24}. One of our 
aims here is to take the sequence of tower of theories to a new level but also 
provide the corresponding explicit perturbative and large $N$ results to 
establish that our underlying universal theory is correct.

Before we proceed to that level of verification we need to first construct the
relevant Lagrangians which populate the tower along a common Wilson-Fisher
fixed point thread in $d$-dimensions. One way to proceed is to do this for a
sequence of critical dimensions by using the universal interaction as the basis
for defining the canonical dimensions of the fields for a specific $D$ and then
construct the spectator part of the Lagrangian which ensures renormalizability.
This will systematically build a tower in much the same way as the ultraviolet 
completion of $O(N)$ $\phi^4$ theory to six dimensions, \cite{15,16}, and 
higher, \cite{17}. However, we have chosen to begin at another point which is 
within the universal theory itself but at the critical point. For the 
established tower (\ref{lag44n}) we noted that the canonical dimensions of the 
fields were determined by dimensionally analysing the kinetic term for $\phi^i$
and the universal interaction to produce the full dimensions of the fields 
which are $\alpha$ and $\beta$ and are given by (\ref{albet}). They satisfy
\begin{equation}
2 \alpha ~+~ \beta ~=~ d ~-~ \chi ~.
\label{albetrel}
\end{equation}
However, this is not the only way to consider the dimensional analysis within
the universal theory. Instead of using the kinetic term for $\phi^i$ and
(\ref{albetrel}) to find (\ref{albet}) we can write down a sequence of 
solutions to (\ref{albetrel}) which includes (\ref{albet}) as a specific case. 
This solution is
\begin{equation}
\alpha ~=~ \mu ~-~ n ~+~ \half \eta ~~~,~~~ \beta ~=~ 2n ~-~ \eta ~-~ \chi
\label{albetn}
\end{equation}
where $n$ is any positive integer. Viewed this way means that one opens up 
new threads of towers since the kinetic term for $\phi^i$ will involve higher
derivatives. Although unlike the scalar theories considered in \cite{29,30,31},
for example, these will be interacting theories with a conformal symmetry at a 
fixed point. More specifically the first few Lagrangians representing the
universal theories will be 
\begin{equation}
\bar{L}^{(8)} ~=~ \frac{1}{2} \left( \Box \phi^i \right)^2 ~+~
\frac{1}{2} \bar{\sigma} \phi^i \phi^i ~-~ \frac{1}{2g_1^2} \bar{\sigma}^2
\label{lag88n}
\end{equation}
and
\begin{equation}
\bar{L}^{(12)} ~=~ \frac{1}{2} \left( \Box \partial_\mu \phi^i \right)^2 ~+~
\frac{1}{2} \bar{\sigma} \phi^i \phi^i ~-~ \frac{1}{2g_1^2} \bar{\sigma}^2
\label{lag1212n}
\end{equation}
for $n$~$=$~$2$ and $3$ in the formulation with the coupling constant 
adjustment used for the large $N$ expansion. Eliminating the auxiliary $\sigma$
field produces
\begin{equation}
L^{(8)} ~=~ \frac{1}{2} \left( \Box \phi^i \right)^2 ~+~
\frac{g_1^2}{8} \left( \phi^i \phi^i \right)^2 
\label{lag8}
\end{equation}
and 
\begin{equation}
L^{(12)} ~=~ \frac{1}{2} \left( \Box \partial_\mu \phi^i \right)^2 ~+~
\frac{g_1^2}{8} \left( \phi^i \phi^i \right)^2 
\label{lag12}
\end{equation}
which are the higher derivative extensions of scalar $O(N)$ $\phi^4$ theory. We
have dropped the overline on the Lagrangian since the $\bar{\sigma}$ field is
absent. Our notation is that the Lagrangian $L^{(D)}$ is perturbatively
renormalizable in $D$ dimensions with critical dimension $4n$ in terms of the
solution (\ref{albetn}). So the critical dimension for the $n$~$=$~$2$ thread
is $8$ and for $n$~$=$~$3$ it is $12$. This explains our notation for the usual
quartic $O(N)$ scalar theory of (\ref{lag4}). As an aside one can in principle 
have a solution with $n$~$=$~$0$ which would have the formal Lagrangian 
\begin{equation}
L^{(0)} ~=~ \frac{1}{2} \phi^i \phi^i ~+~
\frac{g_1^2}{8} \left( \phi^i \phi^i \right)^2 ~. 
\end{equation}
However, this is in effect trivial as there are no spacetime variables to
provide arguments for the fields. At best one is counting graphs with a quartic
interaction. 

\sect{Large $N$ exponents.}

Having introduced new sets of quartic scalar theories with a critical dimension
$D$~$=$~$4n$ and an $O(N)$ symmetry it is possible to determine the various
critical exponents of each theory in the same way that the $n$~$=$~$1$
exponents are known, \cite{21,22,23}. Indeed it turns out that the leading
order exponents for the fields as well as that for $\eta_2$ can be immediately
deduced from \cite{21,22} for any positive value of $n$. This is because in the
construction of Vasil'ev et al the $O(1/N^2)$ diagrams contributing to $\eta_2$
were computed as functions of $\alpha$ and $\beta$ using only (\ref{albetrel})
in the derivation before (\ref{albetn}) was substituted in the final step to 
find the values of the exponents for the $n$~$=$~$1$ case of interest. One
aspect of the evaluation of the $O(1/N^2)$ graphs in \cite{21,22} was the use
of what is termed uniqueness or conformal integration. This was originally
introduced in \cite{44} in strictly three dimensions and then extended by 
others to $d$-dimensions. In the coordinate space formulation of \cite{21,22} 
the rule in brief is that if the sum of the exponents of the lines joining a 
$3$-point vertex is equal to the spacetime dimension then the integral over the
vertex location can be performed. In the large $N$ context for $n$~$=$~$1$ this
was exploited in \cite{21,22}. However, since the canonical dimensions of our 
fields for our higher $n$ solutions satisfy the same uniqueness condition 
independently of $n$ then the use of uniqueness for general $\alpha$ and 
$\beta$ in the derivation of the $O(1/N^2)$ exponents can be simply used for 
$n$~$=$~$2$ and $3$. Therefore, we have revisited \cite{22} and determined 
$\eta_1$, $\eta_2$ and $\chi_1$ for these specific values as examples. These 
will be needed for later in order to allow us to check off explicit 
perturbative results in the respective tower of theories which have each of 
$L^{(8)}$ and $L^{(12)}$ as the seed Lagrangians. It is worth contrasting the 
use of uniqueness here with another aspect of the conformal integration rule. 
This is that there is not one condition for a coordinate space vertex to be 
integrable. If the sum of the exponents at a vertex sum to the spacetime 
dimension plus a positive integer then the vertex can be integrated. See 
\cite{45}, for example, for lectures on this construction. Although the 
resulting expression may be cumbersome. While it is possible to consider 
theories based on the one step from uniqueness criterion it is not our main 
focus here. 

For $L^{(8)}$ the exponents which determine the leading order behaviour of 
scaling behaviour of the respective fields are 
\noindent
\begin{eqnarray}
\eta^{(8)}_1 &=& -~ \frac{ 6 [\mu-1] [\mu-4] \Gamma(2 \mu-3)}
{\Gamma^2(\mu+2) \Gamma(\mu) \Gamma(-1-\mu)} \nonumber \\
\chi^{(8)}_1 &=& -~ \frac{\mu[\mu+1][4\mu^2-30\mu+47]}{9[\mu-3][\mu-4]}
\eta^{(8)}_1 
\label{expd81}
\end{eqnarray}
where we use the notation $\eta^{(D)}_i$ for exponents so as to be clear that 
they are derived from $L^{(D)}$. Extending the formalism of \cite{21,22} to the
next order we find
\noindent
\begin{eqnarray}
\eta^{(8)}_2 &=& \left[ \,-~ \frac{[ 2 \mu^4 - 13 \mu^3 - 2 \mu^2 + 85 \mu 
- 108 ]}{9[\mu - 3][\mu - 4]} \left[ B(3-\mu) - B(\mu-1) \right] \right. 
\nonumber \\ 
&& \left. ~+ \left[ 4 \mu^{10} - 72 \mu^9 + 433 \mu^8 - 697 \mu^7 
- 3085 \mu^6 + 15845 \mu^5 \right. \right. \nonumber \\
&& \left. ~~~~~~ \left. -~ 26504 \mu^4 + 11816 \mu^3 + 15436 \mu^2 
- 16416 \mu + 2592 \right] \right. \nonumber \\
&& \left. ~~~~~ \left/ \left[
18[\mu + 1][\mu - 1][\mu - 2][\mu - 3]^2[\mu - 4]^2\mu \right. \right] 
\frac{}{} \!\!\! \right] {\eta^{(8)}_1 }^2
\label{expd82}
\end{eqnarray}
where 
\begin{equation}
B(z) ~=~ \psi(z) ~+~ \psi(\mu-z)
\end{equation} 
and $\psi(z)$~$=$~$\frac{d\ln\Gamma(z)}{dz}$. Compared to the same exponents
for the $n$~$=$~$1$ case these expressions are more involved. This is because
in the derivation the arguments of the $\Gamma$- and $\psi$-functions will
involve $n$. This is more apparent in the $n$~$=$~$3$ case since we have 
\begin{eqnarray}
\eta^{(12)}_1 &=& -~ \frac{ 80 [\mu-1] [\mu-2] [\mu-6] \Gamma(2 \mu-5)}
{\Gamma^2(\mu+3) \Gamma(\mu) \Gamma(-2-\mu)} \nonumber \\
\chi^{(12)}_1 &=& -~ \frac{\mu[\mu+1][\mu+2][4\mu^4-92\mu^3+767\mu^2-2722\mu
+3453]}{150[\mu-4][\mu-5]^2[\mu-6]} \eta^{(12)}_1  
\label{expd121}
\end{eqnarray}
at leading order in $1/N$ and 
\begin{eqnarray}
\eta^{(12)}_2 &=& \left[ -~ \frac{[ 4 \mu^7 - 80 \mu^6 + 499 \mu^5 - 890 \mu^4 
+ 2791 \mu^3 - 39980 \mu^2 + 153756 \mu - 180000 ]}
{300[\mu - 4][\mu - 5]^2[\mu - 6]} \right. \nonumber \\
&& \left. ~~~~~ \times \left[ B(6-\mu) - B(\mu-3) \right]
\right. \nonumber \\
&& \left. ~- \left[ 24 \mu^{14} - 648 \mu^{13} + 6813 \mu^{12} - 38367 \mu^{11}
+ 197774 \mu^{10} - 1486770 \mu^9 \right. \right. \nonumber \\
&& \left. ~~~~~ + 9018717 \mu^8 - 29983215 \mu^7
+ 37756752 \mu^6 + 56662008 \mu^5 - 238972400 \mu^4 
\right. \nonumber \\
&& \left. ~~~~~ \left. +~ 227278992 \mu^3 + 47272320 \mu^2 
- 150912000 \mu + 34560000 \right] \right. \nonumber \\
&& \left. ~~~~~ \left/ 
\left[ 400[\mu + 2][\mu + 1][\mu - 1][\mu - 2][\mu - 3][\mu - 4]^2
[\mu - 5]^2[\mu - 6]^2\mu \right. \right] \frac{}{} \!\!\! \right] 
{ \eta^{(12)}_1 }^2
\label{expd122}
\end{eqnarray}
at next order. We have included these exponents together with the 
renormalization group functions which are discussed later in an electronic
format in an attached data file. The situation for $n$~$=$~$0$ is different in 
that the leading order large $N$ equations of \cite{21,22} involving $\alpha$ 
and $\beta$ do not have a solution for this value of $n$. The first solution of 
substance is when $n$~$=$~$1$.

Having derived these from the large $N$ form of $L^{(8)}$ and $L^{(12)}$ they 
remain to be checked against explicit perturbative computations of the
underlying renormalization group functions. For both theories we have computed
the anomalous dimensions of $\phi^i$ together with the mass of $\phi^i$ and
the $\beta$-function to the first two terms. This means two loops for the last
two quantities but three loop for the field anomalous dimension. In a quartic
theory there is no one loop contribution to the wave function renormalization
and we require a non-trivial check on our large $N$ exponents. The 
$\beta$-function is key to determining the location of the Wilson-Fisher
fixed points in the respective $\epsilon$ expansions around 
$d$~$=$~$D$~$-$~$2\epsilon$. To determine the renormalization group functions
for the purely quartic theories we used conventional perturbation theory.
However this requires a modification to the treatment of the canonical four 
dimensional quartic scalar theory. Specifically the propagators of $L^{(8)}$ 
and $L^{(12)}$ are $1/(k^2)^2$ and $1/(k^2)^3$ respectively. For the 
renormalization of the $2$-point function we used the integration by parts
algorithm of Laporta, \cite{46}, to reduce the Feynman graphs to a few basic 
master integrals which can then be evaluated by direct methods. To three loops 
the majority of these are simple integrals built out of basic bubbles which are
easy to compute. Two three loop master integrals do not fall into this class
but we have determined their $\epsilon$ expansions with respect to the critical
dimension of each theory by applying the formalism developed by Tarasov in
\cite{47,48}. This allows one to connect the value of such master integrals in 
a dimension $d$ say with that in dimension $(d+2)$ as well as a linear
combination of integrals with fewer propagators. Since the two three loop 
masters are known in four dimensions, \cite{49}, then applying the Tarasov 
method in hand with the Laporta algorithm recursively allowed us to construct 
the required $2$-point three loop masters in eight and twelve dimensions. For 
the $\beta$-function we used the same general approach but employed the vacuum 
bubble expansion method of \cite{50,51}. The corresponding two loop master 
vacuum bubble integrals were similarly deduced using Tarasov's method where the
four dimensional two loop master of \cite{52} was used as the base for 
extending to eight and twelve dimensions. It is worth addressing a concern that
with the higher order propagator in each of these dimensions there may be 
infrared divergences. This is not the case. A propagator such as $1/(k^2)^2$ 
would be infrared problematic in four dimensions but not in six or higher 
dimensions. Equally integrals which involve $1/(k^2)^4$ would not be infrared 
safe in eight dimensions but in twelve dimensions the poles in $\epsilon$ 
emerging from an integral will represent purely ultraviolet divergences. In 
general a propagator of the form $1/(k^2)^\alpha$ is infrared safe in 
$d$~$>$~$2\alpha$ dimensions. Finally, for the renormalization of (\ref{lag8}) 
and (\ref{lag12}) as well as the other theories we examined here we used 
{\sc Qgraf}, \cite{53}, to generate the Feynman diagrams in electronic form. 
This is necessary as we have designed an automatic Feynman graph routine which 
is written in the symbolic manipulation language {\sc Form} and {\sc Tform}, 
\cite{54,55}. The necessary integration by parts relations to obtain the master
integrals were generated via the early and recent versions of the {\sc Reduze} 
package, \cite{56,57}. These relations were encoded in a {\sc Form} module to 
allow us simply to import the Feynman rules module corresponding to whichever 
theory we are interested in.

Consequently we find the first two terms of the renormalization group functions
are 
\begin{eqnarray}
\gamma_\phi^{(8)}(g_1) &=& -~ \frac{[N+2]}{4320} g_1^4 ~+~ 
\frac{[N+2][N+8]}{9331200} g_1^6 ~+~ O(g_1^8) \nonumber \\ 
\gamma_m^{(8)}(g_1) &=& -~ \frac{[N+2]}{36} g_1^2 ~-~ 
\frac{7[N+2]}{12960} g_1^4 ~+~ O(g_1^6) \nonumber \\ 
\beta^{(8)}(g_1) &=& \frac{[N+8]}{36} g_1^4 ~+~ 
\frac{[41N+202]}{19440} g_1^6 ~+~ O(g_1^8) 
\label{rge8}
\end{eqnarray}
for (\ref{lag8}) and
\begin{eqnarray}
\gamma_\phi^{(12)}(g_1) &=& -~ \frac{[N+2]}{4354560} g_1^4 ~-~ 
\frac{293[N+2][N+8]}{2633637888000} g_1^6 ~+~ O(g_1^8) \nonumber \\ 
\gamma_m^{(12)}(g_1) &=& -~ \frac{[N+2]}{720} g_1^2 ~-~ 
\frac{29[N+2]}{3402000} g_1^4 ~+~ O(g_1^6) \nonumber \\ 
\beta^{(12)}(g_1) &=& \frac{[N+8]}{720} g_1^4 ~+~ 
\frac{[255N+1112]}{9072000} g_1^6 ~+~ O(g_1^8) 
\label{rge12}
\end{eqnarray}
for the twelve dimensional case. Structurally these are similar to their four
dimensional counterparts from the point of view of the factor $(N+2)$ and
$(N+8)$. More interestingly neither is asymptotically free in parallel with the
$D$~$=$~$4$ case. However the more intriguing aspect of the results in this
section is that when one computes the respective critical exponents from
(\ref{rge8}) and (\ref{rge12}) near their critical dimensions the coefficients
in the $\epsilon$ expansion are in one-to-one agreement with the same powers
of $\epsilon$ and $1/N$ of $\eta$ and $\chi$. As the $\sigma$ field couples to
the operator $\phi^i \phi^i$ its anomalous dimension in the large $N$ 
formalism is $\eta$~$+$~$\chi$ and the renormalization group function this
corresponds to is $\gamma^{(D)}_m(g_1)$. This agreement between the large $N$
exponents derived from the universal theory residing at the respective eight
and twelve dimensional Wilson-Fisher fixed points and their explicit high
order perturbative renormalization group functions is in keeping with the
situation for $n$~$=$~$1$. Moreover, it has in effect opened up more avenues
to explore aspects of conformal theories in higher dimensions. In principal
one can examine the generalization
\begin{equation}
L^{(4n)} ~=~ \frac{1}{2} \left( \partial_{\mu_1} \ldots \partial_{\mu_n} \phi^i
\right) \left( \partial^{\mu_1} \ldots \partial^{\mu_n} \phi^i \right) ~+~
\frac{1}{8} g_1^2 \left( \phi^i \phi^i \right)^2  
\label{lag4n}
\end{equation}
explicitly. 

We have chosen to stop at twelve dimensions for the practical and not
conceptual reason that there is a limitation to the size of the {\sc Reduze} 
database we constructed using the Laporta algorithm. There is no technical 
obstruction to proceeding further in principle. Instead the direction we take 
is to examine (\ref{expd81}), (\ref{expd82}), (\ref{expd121}) and 
(\ref{expd122}) in dimensions beyond the critical dimensions of each of the 
Lagrangians we have considered in order to make connection with the tower of 
theories which we aim to construct and verify that they lie in the same 
Wilson-Fisher universality class. For instance, if one expands (\ref{expd81}) 
and (\ref{expd82}) in $d$~$=$~$10$~$-$~$2\epsilon$ then unlike in eight 
dimensions both $\eta_1$ and $\eta_2$ begin with $O(\epsilon)$ terms. So unlike
the connection of these exponents with a quartic eight dimensional scalar 
theory where the exponents begin with $O(\epsilon^2)$ the ultraviolet 
completion of $L^{(8)}$ ought to be a cubic theory. This is because a cubic
theory will have a one loop self-energy graph unlike a purely quartic theory.
Equally expanding the exponents around dimensions $12$, $14$ and higher 
dimensions the leading order terms are always $O(\epsilon)$ indicating the 
subsequent completions involve a basic cubic interaction. This is, of course, 
the seed interaction of (\ref{lag88n}). The position with (\ref{expd121}) and 
(\ref{expd122}) is completely similar in terms of the leading order terms in 
$\epsilon$. 

\sect{Ultraviolet complete Lagrangians.}

In order to construct the ultraviolet completions or tower of theories based 
respectively on $L^{(8)}$ and $L^{(12)}$ it is instructive to review the 
position with the conventional Wilson-Fisher universality class. This is also
to allow us to compare structures with the other towers we will construct. The 
key is the use of the canonical dimension of the two basic fields and the 
spacetime dimension the Lagrangian is to be completed in. In the large $N$
expansion the canonical dimensions are necessarily dimension dependent as the 
universal theory is spacetime transcendent. For the theories which are 
renormalizable in a fixed (even) dimension one has to use the canonical 
dimensions for that specific dimension. So when $n$~$=$~$1$ $\sigma$ has 
dimension $2$ and $\phi^i$ has dimension $1$, $2$, $3$ and $4$ in the even 
dimensions between four and ten. One consequence is that in each of these 
dimensions the $\sigma \phi^i \phi^i$ operator dimension is preserved and 
moreover no new $\phi^i$-$\sigma$ interactions can be included. Instead in 
order to ensure renormalizability in each dimension extra pure $\sigma$ 
(spectator) interactions have to be added which can include derivative 
interactions. Given this reasoning we find the following higher dimensional 
extensions of 
$L^{(4)}$, (\ref{lag44}), 
\begin{eqnarray}
L^{(4,6)} &=& \frac{1}{2} \left( \partial_\mu \phi^i \right)^2 ~+~ 
\frac{1}{2} \left( \partial_\mu \sigma \right)^2 ~+~
\frac{g_1}{2} \sigma \phi^i \phi^i ~+~ \frac{g_2}{6} \sigma^3 \nonumber \\
L_\phi^{(4,8)} &=& \frac{1}{2} \partial_\mu \phi^i \partial^\mu \phi^i ~+~
\frac{1}{2} \left( \Box \sigma \right)^2 ~+~
\frac{1}{2} g_1 \sigma \phi^i \phi^i ~+~
\frac{1}{6} g_2 \sigma^2 \Box \sigma ~+~ \frac{1}{24} g_3^2 \sigma^4 
\nonumber \\
L_\phi^{(4,10)} &=& \frac{1}{2} \partial_\mu \phi^i \partial^\mu \phi^i ~+~
\frac{1}{2} \left( \Box \partial^\mu \sigma \right)
\left( \Box \partial_\mu \sigma \right) ~+~
\frac{1}{2} g_1 \sigma \phi^i \phi^i ~+~
\frac{1}{6} g_2 \sigma^2 \Box^2 \sigma \nonumber \\
&& +~ \frac{1}{2} g_3 \sigma \left( \Box \sigma \right)^2 ~+~
\frac{1}{24} g_4^2 \sigma^3 \Box \sigma ~+~ \frac{1}{120} g_5^3 \sigma^5 ~.
\end{eqnarray}
Our notation of $L^{(d_1,d_2)}$ for these Lagrangians is to indicate the
dimension of the base quartic theory, which is $d_1$, and the particular
critical dimension, $d_2$, where it is renormalizable. This is to avoid 
confusion with the same constructions for $L^{(8)}$ and $L^{(12)}$. Given the 
structure of these Lagrangians we could equally well have labelled or 
classified them by the number of derivatives in the $2$-point terms. However, 
we chose the former syntax as we wish to indicate the tower aspect of the 
construction. For each of these theories bar $L^{(4,10)}$ the respective 
renormalization group functions evaluated at the Wilson-Fisher fixed point and 
compared with the corresponding critical exponents of \cite{21,22,23,24} are in
full agreement to the orders they have been computed in perturbation theory. 
For $L^{(4,6)}$ this is to four loops, \cite{15,16,17}, based on the pioneering
work of \cite{18,19}. The verification of $L^{(4,8)}$ was carried out in 
\cite{20} to two loop order. That for $L^{(4,10)}$ will be considered here to 
add confidence in the overall vision of an underlying universal theory in 
$d$-dimensions.

Again when the $n$~$=$~$1$ construction is viewed in this light it is 
straightforward to write down the Lagrangians in the tower with $L^{(8)}$ as
the base theory. In this instance $\sigma$ now has dimension $4$ and $\phi^i$
has dimension $1$, $2$, $3$ and $4$ in eight, ten, twelve and fourteen
dimensions. The seed interaction again is the only one between $\phi^i$ and
$\sigma$ and ensuring the renormalizability is achieved by the spectator
interactions which are purely $\sigma$ dependent. Consequently we have  
\begin{eqnarray}
L^{(8,10)} &=& \frac{1}{2} \left( \Box \phi^i \right)^2 ~+~
\frac{1}{2} \partial_\mu \sigma \partial^\mu \sigma ~+~
\frac{g_1}{2} \sigma \phi^i \phi^i \nonumber \\ 
L^{(8,12)} &=& \frac{1}{2} \left( \Box \phi^i \right)^2 ~+~
\frac{1}{2} \left( \Box \sigma \right)^2 ~+~
\frac{g_1}{2} \sigma \phi^i \phi^i ~+~ \frac{g_2}{6} \sigma^3 \nonumber \\
L^{(8,14)} &=& \frac{1}{2} \left( \Box \phi^i \right)^2 ~+~
\frac{1}{2} \left( \Box \partial^\mu \sigma \right)^2 ~+~
\frac{g_1}{2} \sigma \phi^i \phi^i ~+~ \frac{g_2}{6} \sigma^2 \Box \sigma
\nonumber \\
L^{(8,16)} &=& \frac{1}{2} \left( \Box \phi^i \right)^2 ~+~
\frac{1}{2} \left( \Box^2 \sigma \right)^2 ~+~
\frac{g_1}{2} \sigma \phi^i \phi^i ~+~ \frac{g_2}{6} \sigma^2 \Box^2 \sigma 
\nonumber \\
&& +~ \frac{g_3}{2} \sigma \left( \Box \sigma \right)^2 ~+~ 
\frac{g_4^2}{24} \sigma^4 
\end{eqnarray}
where only independent derivative interactions are included. Repeating the 
exercise for $L^{(12)}$ is similarly straightforward but in this case $\sigma$ 
has canonical dimension $6$. Up to eighteen dimensions we have 
\begin{eqnarray}
L^{(12,14)} &=& \frac{1}{2} \left( \Box \partial_\mu \phi^i \right)^2 ~+~
\frac{1}{2} \partial_\mu \sigma \partial^\mu \sigma ~+~
\frac{g_1}{2} \sigma \phi^i \phi^i \nonumber \\ 
L^{(12,16)} &=& \frac{1}{2} \left( \Box \partial_\mu \phi^i \right)^2 ~+~
\frac{1}{2} \left( \Box \sigma \right)^2  ~+~
\frac{g_1}{2} \sigma \phi^i \phi^i \nonumber \\ 
L^{(12,18)} &=& \frac{1}{2} \left( \Box \partial_\mu \phi^i \right)^2 ~+~
\frac{1}{2} \left( \Box \partial_\mu \sigma \right)^2  ~+~
\frac{g_1}{2} \sigma \phi^i \phi^i ~+~ \frac{g_2}{6} \sigma^3 ~. 
\end{eqnarray}
One of the reasons why we have included a range of Lagrangians built from the
various base Lagrangians is to compare and contrast structural similarities.
For instance the spectator Lagrangians of $L^{(4,6)}$, $L^{(8,12)}$ and
$L^{(12,18)}$ are formally the same although the canonical dimension of the 
$\sigma$ field is not the same in each case. This will generalize to the 
sequence $L^{(4n,6n)}$ but in the dimensions between $4n$ and $6n$ there are
no spectator interactions only a change in the $\sigma$ $2$-point term. A
similar situation arises for the derivative cubic and higher $\sigma$ 
interactions as one proceeds up each tower. It is worth stressing at this stage
that we have merely constructed sequences of higher dimensional renormalizable 
interacting Lagrangians founded on a quartic scalar theory with a higher 
derivative kinetic term. We now need to make the connection with our large $N$ 
exponents in order to extend the Wilson-Fisher threads in this new context.  

\sect{Perturbative results.}

We devote this section to computing the renormalization group functions to as
high a loop order as is calculationally viable for the Lagrangians we have 
constructed in section $4$. To be consistent with other work on the $L^{(4)}$
thread we will use the same convention and notation as that used in \cite{17}. 
In addition we will use the same underlying computational technology described 
in \cite{17}. For instance, an efficient algorithm was used to easily access 
the renormalization of $3$-point vertices. This exploited the fact that a 
propagator $1/(k^2)^\alpha$ is infrared safe in $d$~$>$~$2\alpha$ dimensions to
allow us to simply make the replacement 
\begin{equation}
\frac{1}{(k^2)^p} ~\mapsto~ \frac{1}{(k^2)^p} ~+~ \frac{\xi g_i}{(k^2)^{p+1}}
\end{equation}
for either a $\phi^i$ or $\sigma$ propagator where $g_i$ is the appropriate
coupling constant and integer $p$. The parameter $\xi$ is used to limit the 
expansion as one only requires a single insertion on a propagator to generate 
vertex graphs from the $2$-point functions. As noted in \cite{17} this approach 
is only applicable in the Lagrangians with $3$-point vertices. Those 
Lagrangians with quartic and higher spectator interactions require a more 
direct approach such as that used for $L^{(4,8)}$ in \cite{20}. For theories
with such interactions we will use the same method. First, the renormalization
of $2$-point Green's functions will proceed as just described but the 
propagator shift is not included as it would omit graphs with quartic and 
higher interactions. In this case the $3$-point vertex functions are 
renormalized by considering the Green's function at either a completely 
symmetric point or at a completely off-shell point. The former is appropriate 
to use when there is either non-derivative $3$-point interactions or a single 
$3$-point vertex. The off-shell configuration is used when there is more than 
one $3$-point interaction and they involve derivative couplings as in the case 
of $L^{(4,10)}$. Equally the $4$-point functions are treated by evaluating the
vertex function at the completely symmetric point. Further background and
calculational detail for both these instances can be found in \cite{20}. 
Finally for the higher $n$-point functions the vacuum bubble method already 
outlined for the base $\phi^4$ theories was used. In terms of loop orders 
the theories in the higher dimensions were renormalized mostly to two loops
but to three loops for a few cases above the critical dimension of the base
Lagrangian. This is because there are practical limitations in the construction
of the databases we used to apply the Laporta and Tarasov algorithms, 
\cite{46,47,48}. The increase in the powers of the propagators means that to 
build the three loop $2$-point masters beyond twelve dimensions, which requires
a significant amount of integration by parts even for non-tensor integrals, was
not viable. However, we take the point of view that it will be evident even 
with two loop renormalization group functions that the connection between all 
the theories will be established. 
 
First, we record our results for the theories along the thread based on
$L^{(8)}$. We have 
\begin{eqnarray}
\gamma^{(8,10)}_\phi (g_1) &=&
\frac{g_1^2}{120} ~+~ [ 194 N - 567 ] \frac{g_1^4}{864000} \nonumber \\
&& +~ [ - 37786 N^2 - 259420 N + 648000 \zeta_3 + 505299 ] 
\frac{g_1^6}{21772800000} ~+~ O(g_1^8) \nonumber \\
\gamma^{(8,10)}_\sigma (g_1) &=& 
-~ \frac{N g_1^2}{60} ~+~ \frac{167 N g_1^4}{216000} \nonumber \\ 
&& +~ [ 259847 N - 648000 \zeta_3 + 256266 ] \frac{N g_1^6}{10886400000} ~+~
O(g_1^8) \nonumber \\
\beta^{(8,10)} (g_1) &=& [ - N + 6 ] \frac{g_1^3}{240} ~+~ 
[ - 197 N - 297 ] \frac{g_1^5}{288000} \nonumber \\
&& +~ [ - 859789 N^2 + 25272000 \zeta_3 N - 38231814 N - 38232000 \zeta_3 
\nonumber \\
&& ~~~~ +~ 43101039 ] \frac{g_1^7}{43545600000} ~+~ O(g_1^9)
\end{eqnarray} 
for $L^{(8,10)}$. Our three loop results for $L^{(8,12)}$ are  
\begin{eqnarray}
\gamma^{(8,12)}_\phi (g_1,g_2) &=& \frac{g_1^2}{280} ~+~ 
[ - 1587 N g_1^2 - 9334 g_1^2 - 6160 g_1 g_2 - 1587 g_2^2 ]
\frac{g_1^2}{197568000} \nonumber \\
&& +~ [ \, - 13130181 N^2 g_1^4 + 175046616 N g_1^4 + 8890560000 \zeta_3 g_1^4
- 9803169176 g_1^4
\nonumber \\
&& ~~~~
+ 425268522 N g_1^3 g_2 + 712313280 g_1^3 g_2
+ 76042962 N g_1^2 g_2^2 
\nonumber \\
&& ~~~~
+ 8890560000 \zeta_3 g_1^2 g_2^2
- 10046446142 g_1^2 g_2^2 - 150402798 g_1 g_2^3
\nonumber \\
&& ~~~~
+ 209985345 g_2^4 ] \frac{g_1^2}{1254635827200000} ~+~ O(g_i^8) \nonumber \\
\gamma^{(8,12)}_\sigma (g_1,g_2) &=& [N g_1^2 + g_2^2] \frac{1}{560} 
\nonumber \\
&& +~ [ - 6254 N g_1^4 - 6160 N g_1^3 g_2 - 1587 N g_1^2 g_2^2 - 4667 g_2^4 ]
\frac{1}{197568000} \nonumber \\ 
&& +~ [ 930040938 N^2 g_1^6 + 8890560000 \zeta_3 N g_1^6 - 9419379728 N g_1^6
\nonumber \\
&& ~~~
+ 194624640 N^2 g_1^5 g_2 + 491748768 N g_1^5 g_2
- 34230843 N^2 g_1^4 g_2^2 
\nonumber \\
&& ~~~
+ 22226400000 \zeta_3 N g_1^4 g_2^2
- 26093250026 N g_1^4 g_2^2 + 1287984600 N g_1^3 g_2^3
\nonumber \\
&& ~~~
+ 417101400 N g_1^2 g_2^4 + 4445280000 \zeta_3 g_2^6
\nonumber \\
&& ~~~
- 4603622893 g_2^6 ] \frac{1}{2509271654400000} ~+~ O(g_i^8) \nonumber \\
\beta^{(8,12)}_1 (g_1,g_2) &=& [ 3 N g_1^2 + 40 g_1^2 + 28 g_1 g_2 + 3 g_2^2 ] 
\frac{g_1}{3360} \nonumber \\
&& +~ [ 12042 N g_1^4 - 53464 g_1^4 - 133308 N g_1^3 g_2 + 490392 g_1^3 g_2 
- 14283 N g_1^2 g_2^2 
\nonumber \\
&& ~~~ + 66956 g_1^2 g_2^2 + 57960 g_1 g_2^3 - 42003 g_2^4 ]
\frac{g_1}{3556224000} \nonumber \\ 
&& +~ [ \, - 14938245342 N^2 g_1^6 - 617004864000 \zeta_3 N g_1^6
+ 3109795833456 N g_1^6 
\nonumber \\
&& ~~~~
+ 2311545600000 \zeta_3 g_1^6 + 1737998549536 g_1^6
- 2875390812 N^2 g_1^5 g_2 
\nonumber \\
&& ~~~~
+ 1867017600000 \zeta_3 N g_1^5 g_2
- 2203538203104 N g_1^5 g_2 
\nonumber \\
&& ~~~~
+ 547658496000 \zeta_3 g_1^5 g_2
- 423865546832 g_1^5 g_2 - 308077587 N^2 g_1^4 g_2^2
\nonumber \\
&& ~~~~
+ 200037600000 \zeta_3 N g_1^4 g_2^2 - 178146641322 N g_1^4 g_2^2
+ 2261758464000 \zeta_3 g_1^4 g_2^2 
\nonumber \\
&& ~~~~
+ 6401866158256 g_1^4 g_2^2
+ 8781904656 N g_1^3 g_2^3 + 497871360000 \zeta_3 g_1^3 g_2^3
\nonumber \\
&& ~~~~
- 586662087088 g_1^3 g_2^3 + 3753912600 N g_1^2 g_2^4
+ 49787136000 \zeta_3 g_1^2 g_2^4 
\nonumber \\
&& ~~~~
+ 2136750334680 g_1^2 g_2^4
+ 373403520000 \zeta_3 g_1 g_2^5 - 374235507660 g_1 g_2^5
\nonumber \\
&& ~~~~
+ 40007520000 \zeta_3 g_2^6 
- 41432606037 g_2^6 ] \frac{g_1}{45166889779200000} ~+~ O(g_i^9) \nonumber \\
\beta^{(8,12)}_2 (g_1,g_2) &=& [ 28 N g_1^3 + 9 N g_1^2 g_2 + 37 g_2^3 ] 
\frac{1}{3360} \nonumber \\
&& +~ [ 173880 N g_1^5 + 32826 N g_1^4 g_2 + 241164 N g_1^3 g_2^2 
\nonumber \\
&& ~~~ - 159651 N g_1^2 g_2^3 + 96073 g_2^5 ] \frac{1}{3556224000} 
\nonumber \\
&& +~ [ 121231189296 N^2 g_1^7 + 1194891264000 \zeta_3 N g_1^7
- 1408785519120 N g_1^7 
\nonumber \\
&& ~~~
- 4290992658 N^2 g_1^6 g_2
+ 1434936384000 \zeta_3 N g_1^6 g_2 + 7192728107344 N g_1^6 g_2
\nonumber \\
&& ~~~
- 90566288568 N^2 g_1^5 g_2^2 + 1344252672000 \zeta_3 N g_1^5 g_2^2
- 1313378645040 N g_1^5 g_2^2 
\nonumber \\
&& ~~~
- 10178257905 N^2 g_1^4 g_2^3
+ 1869684768000 \zeta_3 N g_1^4 g_2^3 + 4814949206106 N g_1^4 g_2^3
\nonumber \\
&& ~~~
+ 746807040000 \zeta_3 N g_1^3 g_2^4 - 890895567408 N g_1^3 g_2^4
- 34486893072 N g_1^2 g_2^5 
\nonumber \\
&& ~~~
+ 941510304000 \zeta_3 g_2^7 + 1196618048425 g_2^7 ] 
\frac{1}{45166889779200000} \nonumber \\
&& +~ O(g_i^9) 
\end{eqnarray}
where $\zeta_z$ is the Riemann zeta function. However, for $L^{(8,14)}$ 
computational limitations meant we can only provide two loop results which are 
\begin{eqnarray}
\gamma_\phi^{(8,14)} (g_1,g_2) &=& \frac{g_1^2}{1120} ~+~ 
[ 107964 N g_1^2 - 1533897 g_1^2 + 718200 g_1 g_2 - 54586 g_2^2 ] 
\frac{g_1^2}{768144384000} \nonumber \\
&& +~ O(g_i^6) \nonumber \\
\gamma_\sigma^{(8,14)} (g_1,g_2) &=& [ - 18 N g_1^2 + 7 g_2^2 ] 
\frac{1}{136080} \nonumber \\
&& +~ [ 13056633 N g_1^4 - 6826680 N g_1^3 g_2 + 467334 N g_1^2 g_2^2 
\nonumber \\
&& ~~~~
+~ 275849 g_2^4 ] \frac{1}{23332385664000} ~+~ O(g_i^6) \nonumber \\
\beta_1^{(8,14)}(g_1,g_2) &=& 
[ - 18 N g_1^2 + 621 g_1^2 - 252 g_1 g_2 + 7 g_2^2 ] \frac{g_1}{272160}
\nonumber \\
&& +~ [ 171159480 N g_1^4 - 12056931 g_1^4 - 67296096 N g_1^3 g_2 
- 377785296 g_1^3 g_2 \nonumber \\
&& ~~~~ + 1869336 N g_1^2 g_2^2 - 7019838 g_1^2 g_2^2 
-~ 59274432 g_1 g_2^3 \nonumber \\
&& ~~~~
+~ 1103396 g_2^4 ] \frac{g_1}{186659085312000} ~+~ O(g_i^6) \nonumber \\
\beta_2^{(8,14)}(g_1,g_2) &=& [ 72 N g_1^3 - 6 N g_1^2 g_2 + g_2^3 ] 
\frac{1}{30240} \nonumber \\
&& +~ [ - 7007148 N g_1^5 + 25365069 N g_1^4 g_2 + 27512136 N g_1^3 g_2^2 
\nonumber \\
&& ~~~~
-~ 639018 N g_1^2 g_2^3 + 2198333 g_2^5 ] \frac{1}{15554923776000} ~+~ 
O(g_i^7) ~. 
\end{eqnarray} 
Equally for  similar computational constraints we could only determine the full
renormalization group functions for $L^{(8,16)}$ at one loop. We found 
\begin{eqnarray}
\gamma^{(8,16)}_\phi (g_i) &=& \frac{g_1^2}{6048} \nonumber \\
&& + \left[ - 1468755 N g_1^2 - 55406142 g_1^2 - 4477968 g_1 g_2
- 29420424 g_1 g_3 \right. \nonumber \\
&& \left. ~~~ - 2792128 g_2^2 + 2456232 g_2 g_3 - 2116377 g_3^2 \right]
\frac{g_1^2}{1003811081011200} ~+~ O(g_i^6) \nonumber \\ 
\gamma^{(8,16)}_\sigma (g_i) &=&
\left[ 45 N g_1^2 + 296 g_2^2 - 384 g_2 g_3 + 159 g_3^2 \right]
\frac{1}{5987520} \nonumber \\ 
&& + \left[ - 2880219870 N g_1^4 - 1664530560 N g_1^3 g_2
- 338294880 N g_1^3 g_3 \right. \nonumber \\
&& \left. ~~~ - 479937600 N g_1^2 g_2^2 + 584739000 Ng_1^2 g_2 g_3
- 243713475 N g_1^2 g_3^2 \right. \nonumber \\
&& \left. ~~~ - 1869662576 g_2^4 + 691909088 g_2^3 g_3
- 1637070624 g_2^2 g_3^2 - 2769180480 g_2^2 g_4^2 \right. \nonumber \\
&& \left. ~~~ + 4241645772 g_2 g_3^3
+ 3466964160 g_2 g_3 g_4^2 - 1825036161 g_3^4 - 1227798000 g_3^2 g_4^2
\right. \nonumber \\
&& \left. ~~~ - 86444820 g_4^4 \right] \frac{1}{496886485100544000} ~+~ 
O(g_i^6) \nonumber \\
\beta^{(8,16)}_1 (g_i) &=&
\left[ 45 N g_1^2 + 4356 g_1^2 + 1584 g_1 g_2 + 792 g_1 g_3 + 296 g_2^2
- 384 g_2 g_3 + 159 g_3^2 \right] \frac{g_1}{11975040} \nonumber \\ 
&& +~ O(g_i^4) \nonumber \\
\beta^{(8,16)}_2 (g_i) &=&
\left[ 1782 N g_1^3 + 135 N g_1^2 g_2 + 2516 g_2^3 - 866 g_2^2 g_3
- 447 g_2 g_3^2 + 3168 g_2 g_4^2 - 330 g_3^3 \right. \nonumber \\
&& \left. - 2376 g_3 g_4^2 
\right] \frac{1}{11975040} ~+~ O(g_i^5) \nonumber \\
\beta^{(8,16)}_3 (g_i) &=&
\left[ 2376 N g_1^3 + 135 N g_1^2 g_3 + 176 g_2^3 - 696 g_2^2 g_3
- 1020 g_2 g_3^2 + 961 g_3^3 \right] \frac{1}{11975040} \nonumber \\
&& +~ O(g_i^5) \nonumber \\
\beta^{(8,16)}_4 (g_i) &=&
[ 1782 N g_1^4 + 45 N g_1^2 g_4^2 + 352 g_2^4 + 704 g_2^3 g_3 + 528 g_2^2 g_3^2 + 1880 g_2^2 g_4^2 + 176 g_2 g_3^3 \nonumber \\ 
&& ~ + 1200 g_2 g_3 g_4^2 + 22 g_3^4 + 555 g_3^2 g_4^2 + 891 g_4^4 ]
\frac{1}{2993760} ~+~ O(g_i^6) ~.
\end{eqnarray}
The two loop wave function anomalous dimensions were computed to provide a 
non-trivial check on the one loop coupling constant renormalization. 

For the $n$~$=$~$3$ thread we are limited to two loops throughout but our
renormalization group functions for the first three Lagrangians are
\begin{eqnarray}
\gamma_\phi^{(12,14)}(g_1) &=& -~ \frac{g_1^2}{6048} ~+~
[ - 103653 N + 179983 ] \frac{g_1^4}{460886630400} ~+~ O(g_1^6) \nonumber \\
\gamma_\sigma^{(12,14)}(g_1) &=& -~ \frac{N g_1^2}{1120} ~+~ 
\frac{611 N g_1^4}{4267468800} ~+~ O(g_1^6) \nonumber \\
\beta^{(12,14)}(g_1) &=& [ - 27 N + 74 ] \frac{g_1^3}{120960} ~+~ 
[ - 3489939 N + 6027443 ] \frac{g_1^5}{4608866304000} \nonumber \\
&& +~ O(g_1^7) \nonumber \\
\gamma_\phi^{(12,16)}(g_1) &=& -~ \frac{g_1^2}{15120} ~+~ 
[ 8661 N + 12118 ] \frac{g_1^4}{1152216576000} ~+~ O(g_1^6) \nonumber \\
\gamma_\sigma^{(12,16)}(g_1) &=& \frac{N g_1^2}{10080} ~+~ 
\frac{47 N g_1^4}{96018048000} ~+~ O(g_1^6) \nonumber \\
\beta^{(12,16)}(g_1) &=& [3 N + 8 ] \frac{g_1^3}{120960} ~+~ 
[ - 31845 N + 142192 ] \frac{g_1^5}{2304433152000} ~+~ O(g_1^7)
\end{eqnarray}
and 
\begin{eqnarray}
\gamma_\phi^{(12,18)}(g_1,g_2) &=& -~ \frac{g_1^2}{66528} \nonumber \\
&& +~ [ - 148285 N g_1^2 + 143716 g_1^2 + 440286 g_1 g_2 - 148285 g_2^2 ]
\frac{g_1^2}{1226880210124800} \nonumber \\
&& +~ O(g_i^6) \nonumber \\
\gamma_\sigma^{(12,18)}(g_1,g_2) &=& -~ [ N g_1^2 + g_2^2 ] \frac{1}{133056} 
\nonumber \\ 
&& +~ [ - 76427 N g_1^4 + 440286 N g_1^3 g_2 - 148285 N g_1^2 g_2^2 
\nonumber \\
&& ~~~~ +~ 71858 g_2^4 ] \frac{1}{1226880210124800} ~+~ O(g_i^6) \nonumber \\
\beta_1^{(12,18)}(g_1,g_2) &=& [ - 5 N g_1^2 + 13 g_1^2 + 33 g_1 g_2
- 5 g_2^2 ] \frac{g_1}{1330560} \nonumber \\
&& +~ [ 150438125 N g_1^4 + 590701817 g_1^4 + 97868100 N g_1^3 g_2 
+ 945987075 g_1^3 g_2 \nonumber \\
&& ~~~~ -~ 14828500 N g_1^2 g_2^2 + 290724292 g_1^2 g_2^2 
+ 214657575 g_1 g_2^3 \nonumber \\
&& ~~~~ +~ 7185800 g_2^4 ] \frac{g_1}{245376042024960000} ~+~ O(g_i^7) 
\nonumber \\
\beta_2^{(12,18)}(g_1,g_2) &=& [ 11 N g_1^3 - 5 N g_1^2 g_2 + 6 g_2^3 ] 
\frac{1}{443520} \nonumber \\ 
&& +~ [ 214657575 N g_1^5 + 139088071 N g_1^4 g_2 + 204846675 N g_1^3 g_2^2 
\nonumber \\
&& ~~~~ +~ 12091250 N g_1^2 g_2^3 + 190227857 g_2^5 ] 
\frac{1}{81792014008320000} ~+~ O(g_i^7) \,.
\end{eqnarray}
Finally we note the renormalization group functions for $L^{(4,10)}$ which
extends the $n$~$=$~$1$ thread to the next dimension are
\begin{eqnarray}
\gamma^{(4,10)}_\phi (g_i) &=& -~ \frac{g_1^2}{40} \nonumber \\
&& +~ [ -~ 5301 N g_1^2 + 16758 g_1^2 + 120540 g_1 g_2 + 302820 g_1 g_3
- 14114 g_2^2 - 18032 g_2 g_3 \nonumber \\
&& ~~~~
- 15779 g_3^2 ] \frac{g_1^2}{254016000} ~+~ O(g_i^4) \nonumber \\
\gamma^{(4,10)}_\sigma (g_i) &=&
[ - 9 N g_1^2 - 86 g_2^2 + 112 g_2 g_3 - 71 g_3^2 ] \frac{1}{15120} 
\nonumber \\ 
&& +~ [ 664524 N g_1^4 + 6713280 N g_1^3 g_2 + 1451520 N g_1^3 g_3 
- 1128852 N g_1^2 g_2^2 \nonumber \\
&& ~~~~ + 1202544 N g_1^2 g_2 g_3 - 797022 N g_1^2 g_3^2 + 4415512 g_2^4 
+ 6451480 g_2^3 g_3 \nonumber \\
&& ~~~~ - 14360000 g_2^2 g_3^2 - 3621996 g_2^2 g_4^2 - 10763088 g_2 g_3^3 
+ 10666782 g_2 g_3 g_4^2 \nonumber \\
&& ~~~~ + 8993886 g_3^4 - 1885086 g_3^2 g_4^2 - 496125 g_4^4 ] 
\frac{1}{96018048000} ~+~ O(g_i^4) \nonumber \\
\beta^{(4,10)}_1 (g_i) &=& [ - 9 N g_1^2 + 504 g_1^2 + 840 g_1 g_2 
+ 420 g_1 g_3 - 86 g_2^2 + 112 g_2 g_3 - 71 g_3^2 ] \frac{g_1}{30240} ~+~ 
O(g_i^5) \nonumber \\
\beta^{(4,10)}_2 (g_i) &=&
[ 756 N g_1^3 - 81 N g_1^2 g_2 + 1298 g_2^3 + 1764 g_2^2 g_3 
- 1395 g_2 g_3^2 - 1134 g_2 g_4^2 - 308 g_3^3 \nonumber \\
&& ~ + 1134 g_3 g_4^2 ] \frac{1}{90720} ~+~ O(g_i^5) \nonumber \\
\beta^{(4,10)}_3 (g_i) &=&
[ 756 N g_1^3 - 81 N g_1^2 g_3 - 448 g_2^3 - 1782 g_2^2 g_3 
+ 3024 g_2 g_3^2 + 1134 g_2 g_4^2 + 565 g_3^3 \nonumber \\
&& ~ - 1134 g_3 g_4^2 ] \frac{1}{90720} ~+~ O(g_i^6) \nonumber \\
\beta^{(4,10)}_4 (g_i) &=&
[ 11340 N g_1^4 - 81 N g_1^2 g_4^2 + 896 g_2^4 - 4256 g_2^3 g_3 
- 7728 g_2^2 g_3^2 + 234 g_2^2 g_4^2 \nonumber \\ 
&& ~ - 4088 g_2 g_3^3 + 8820 g_2 g_3 g_4^2 + 2268 g_2 g_5^3 - 700 g_3^4 
+ 3015 g_3^2 g_4^2 - 2268 g_3 g_5^3 \nonumber \\ 
&& ~ - 1134 g_4^4 ] \frac{1}{6804} ~+~ O(g_i^6) \nonumber \\
\beta^{(4,10)}_5 (g_i) &=&
[ - 27216 N g_1^5 - 81 N g_1^2 g_5^3 - 3584 g_2^5 - 8960 g_2^4 g_3 
- 8960 g_2^3 g_3^2 + 10080 g_2^3 g_4^2 \nonumber \\
&& ~ - 4480 g_2^2 g_3^3 + 15120 g_2^2 g_3 g_4^2 - 10854 g_2^2 g_5^3 
- 1120 g_2 g_3^4 + 7560 g_2 g_3^2 g_4^2 \nonumber \\
&& ~ - 9072 g_2 g_3 g_5^3 - 5670 g_2 g_4^4 - 112 g_3^5 + 1260 g_3^3 g_4^2 
- 3159 g_3^2 g_5^3 - 2835 g_3 g_4^4 \nonumber \\
&& ~ + 5670 g_4^2 g_5^3 ] \frac{1}{54432} ~+~ O(g_i^7) ~.
\end{eqnarray}
All the renormalization group functions have been determined using dimensional
regularization with the renormalization constants defined with respect to the
$\MSbar$ scheme. It is worth noting that in the critical dimension of each
Lagrangian we used the coupling constants are dimensionless in that dimension
but the standard arbitrary scale is introduced to preserve dimensionlessness of
the couplings in the regularized theory.  

The main reason for constructing these renormalization group functions is to
verify that the critical exponents at the Wilson-Fisher fixed point are
consistent with the large $N$ critical exponents given in section $3$ for each 
of the underlying universal theories. In order to carry out the comparison we
follow the process given in \cite{15,16} and first find the values of the 
critical coupling constants, $g_{i\,c}$, by solving 
\begin{equation}
\beta_j^{(d_1,d_2)}(g_{i\,c}) ~=~ 0
\end{equation}
where $g_{i\,c}$ is a power series in $1/N$. Each coefficient is itself a power
series in $\epsilon$ aside from the leading order $1/N$ term which only 
involves $\epsilon$ due to the structure of the $N$ dependence at two and 
higher loops. Once these critical couplings are determined the field anomalous
dimensions $\gamma_\phi^{(d_1,d_2)}(g_i)$ and $\gamma_\sigma^{(d_1,d_2)}(g_i)$
are evaluated at criticality as series in $1/N$. Then the coefficients of each
term in $\epsilon$ of each successive power of $1/N$ should be in total 
agreement with the critical exponents $\eta$ and $\eta$~$+$~$\chi$ 
respectively. We have checked this correspondence holds for all of the sets of
renormalization group functions for the threads $n$~$=$~$2$ and $3$ we have 
computed and the large $N$ exponents (\ref{expd81}), (\ref{expd82}), 
(\ref{expd121}) and (\ref{expd122}). Such agreement should be regarded as 
evidence for the underlying universality of the core interaction across the 
dimensions in the same spirit as that of the original and well-established 
universality of the Wilson-Fisher fixed point of $O(N)$ $\phi^4$ theory or 
$n$~$=$~$1$ thread in the present language. Equally the agreement is a 
reassuring check that we have correctly performed the renormalization to 
several loop orders which relied on elevating the various master integrals to 
higher dimensions. There are already several internal checks on the
perturbative results in that the double and triple poles of the two and three
loop renormalization constants are already predetermined by the lower loop
results. We have ensured that these have been satisfied first before recording
our two and three loop expressions. 

Having established the connection with the underlying universal theory it is
worth briefly analysing aspects of the non-trivial fixed point structure of
each theory and in particular the location, if it exists, of the conformal
window. This falls into two classes of analysis. In QCD there is a conformal 
window when the signs of the one and two terms of the strictly four dimensional
$\beta$-function are different with the non-trivial fixed point being called 
the Banks-Zaks fixed point, \cite{58}. Such a class of critical points of the 
renormalization group equations is a feature of single coupling theory. Several
of the theories we have constructed have the same potential property and we 
have determined the conformal windows for these. For instance, the two loop 
term of $\beta^{(8,10)}(g)$ is always negative but the one loop term changes 
sign at $N$~$=$~$6$. Hence it is straightforward to see that the conformal 
window is $1$~$\leq$~$N$~$<$~$6$. When $N$~$=$~$6$ the two and three loop terms
are both negative which is the reason for the strict inequality. Above 
$N$~$=$~$6$ the non-trivial critical coupling of the so-called Banks-Zaks fixed
point is real whereas it becomes pure imaginary below $N$~$=$~$6$. By contrast 
the eight dimensional $\phi^4$ theory is not asymptotically free and the two 
loop term of its $\beta$-function is positive. There is a parallel picture for 
the $n$~$=$~$3$ thread as the base twelve dimensional $\phi^4$ theory has two 
positive $\beta$-function terms but there are Banks-Zaks fixed points for the 
higher dimensional single coupling theories. For instance for the $(12,14)$ 
theory there is a conformal window at $N$~$=$~$1$ and $2$ with real critical 
couplings for $N$~$\geq$~$3$. By contrast $\beta^{(12,16)}(g)$ has a positive 
one loop term but the two loop term is negative for $N$~$\geq$~$5$. So there is
a Banks-Zaks type fixed point for this range of $N$.

The second class of fixed point analysis concerns theories with more than one
coupling constant. To access the conformal window in this instance we have to
solve a set of equations, \cite{15,16}, which for the two coupling theories
considered here, are
\begin{equation}
\beta_1(g_i) ~=~ \beta_2(g_i) ~=~ 0 ~~,~~
\frac{\partial \beta_1}{\partial g_1} \frac{\partial \beta_2}{\partial g_2}
~-~ \frac{\partial \beta_1}{\partial g_2}
\frac{\partial \beta_2}{\partial g_1} ~=~ 0 ~.
\label{hess2}
\end{equation}
The first two equations determine the critical couplings and the final one,
which is the vanishing of the Hessian, provides the condition where there is a 
change in the stability property of a fixed point. Moreover, as in \cite{15,16}
we can determine the window as a perturbative series in $\epsilon$ which, in 
principle, provides insight into other dimensions. For 
$\beta^{(8,12)}_i(g_1,g_2)$ we found four solutions to (\ref{hess2}) for the 
critical value of $N$ which are
\begin{eqnarray}
N^{(8,12)}_{(A)} &=& 1.015123 ~-~ 0.024469 \epsilon ~-~ 
0.324484 \epsilon^2 ~+~ O(\epsilon^3) \nonumber \\
N^{(8,12)}_{(B)} &=& -~ 0.366698 ~+~ 0.451194 \epsilon ~-~ 
41.675880 \epsilon^2 ~+~ O(\epsilon^3) \nonumber \\
N^{(8,12)}_{(C)} &=& -~ 910.687640 ~+~ 2668.861873 \epsilon ~-~ 
1565.439288 \epsilon^2 ~+~ O(\epsilon^3) 
\end{eqnarray}
where solution $B$ has real critical couplings whereas the other two cases are
pure imaginary. Given the non-unitary nature of solution $A$ and the negative
corrections in the $\epsilon$ to the low value for the critical value of $N$
defining the conformal window boundary. So it would appear that for this
theory there is no interesting structure. By contrast for the theory based on
the related group $Sp(N)$ the conformal window is determined from the 
negative solutions, \cite{59}. So that theory would appear to have a conformal
window around $N$~$=$~$910$. A similar feature was observed in the $(4,8)$
case. There were three real solutions for the conformal window boundary for
$\beta^{(8,14)}_i(g_1,g_2)$ which are
\begin{eqnarray}
N^{(8,14)}_{(A)} &=& 602.601144 ~-~ 33341.878584 \epsilon ~+~ O(\epsilon^2) 
\nonumber \\
N^{(8,14)}_{(B)} &=& 0.627879 ~-~ 1.399181 \epsilon ~+~ O(\epsilon^2) 
\nonumber \\
N^{(8,14)}_{(C)} &=& -~ 186.979023 ~+~ 45848.701747 \epsilon ~+~
O(\epsilon^2) ~.
\end{eqnarray}
By contrast there is a clear indication of a conformal window here with a
relatively high value for $N$ similar to the $(4,6)$ theory, \cite{15,16,17}.
The first multicoupling example for the $n$~$=$~$3$ thread occurs for
$\beta^{(12,18)}_i(g_1,g_2)$ and gives 
\begin{eqnarray}
N^{(12,18)}_{(A)} &=& 113.894634 ~-~ 1653.078171 \epsilon ~+~ 
O(\epsilon^2) \nonumber \\
N^{(12,18)}_{(B)} &=& 1.116917 ~-~ 2.093367 \epsilon ~+~ 
O(\epsilon^2) \nonumber \\
N^{(12,18)}_{(C)} &=& -~ 0.032996 ~-~ 0.713266 \epsilon ~+~ O(\epsilon^2) 
\end{eqnarray}
which has parallels with the previous case. For the remaining theories the 
increase in the number of couplings and hence $\beta$-functions together with a
substantial Hessian meant that our computer resources rather than any principle
were not powerful enough to solve the system of equations in general.

\sect{Extensions.}

Our main focus has been on $O(N)$ scalar theories and the new threads of 
theories which follow from new integer solutions to the canonical relation
between the field critical exponents derived from the critical exponent defined
by the force-matter vertex. It transpires that such an exercise is not limited 
to scalar theories. The $O(N)$ symmetric Gross-Neveu (GN) model, \cite{36}, and
the non-abelian Thirring model (NATM), \cite{37,38}, have also been considered 
in the context of the Vasil'ev et al large $N$ expansion, 
\cite{60,61,62,63,64,65,66,67,68}. In the case of the latter theory it is the 
large $\Nf$ expansion, where $\Nf$ is the number of quark flavours, rather than
the number of colours of the non-abelian symmetry which is the expansion 
parameter. Therefore we have constructed several higher derivative extensions 
of each of these base theories in the same spirit as the scalar theories. First
we recall that the Lagrangian of the two dimensional Gross-Neveu model with an 
$SU(N)$ symmetry is, \cite{36}, 
\begin{equation}
L_{\mbox{\footnotesize{GN}}}^{(2)} ~=~ i \bar{\psi}^i \partialslash \psi^i ~+~
\frac{1}{2} g_1^2 \left( \bar{\psi}^i \psi^i \right)^2  
\label{laggn2}
\end{equation}
which like (\ref{lag4}) can be rewritten in terms of an auxiliary field
$\sigma$ to give 
\begin{equation}
L_{\mbox{\footnotesize{GN}}}^{(2)} ~=~ i \bar{\psi}^i \partialslash \psi^i ~+~
\frac{1}{2} g_1 \sigma \bar{\psi}^i \psi^i  ~-~ \frac{1}{2} \sigma^2 ~.
\label{laggn22}
\end{equation}
Equally one can develop a large $N$ expansion for (\ref{laggn22}) using the
same approach as \cite{21,22}. In this instance the respective scaling 
dimensions of $\psi$ and $\sigma$ are $\alpha$ and $\beta$ and are given by  
\begin{equation}
\tilde{\alpha} ~=~ \mu ~+~ \half \eta ~~~,~~~ 
\tilde{\beta} ~=~ 1 ~-~ \eta ~-~ \chi 
\end{equation}
giving 
\begin{equation}
2 \tilde{\alpha} ~+~ \tilde{\beta} ~=~ d ~+~ 1 ~-~ \chi ~.
\label{gnuniq}
\end{equation}
In defining these dimensions we follow \cite{60} and do not include the
dimension deriving from $\partialslash$ in the kinetic term of the Lagrangian. 
This is so that $\tilde{\alpha}$ corresponds to the exponent in the critical 
point large $N$ propagators analogous to (\ref{critprop}). With this convention
for a vertex of the Gross-Neveu form the uniqueness condition for its conformal
integration translates to the sum of the vertex exponents is $(d$~$+$~$1)$,
\cite{60}.

The Gross-Neveu model is therefore parallel to the $O(N)$ scalar theories in 
that the vertices are unique in the absence of the vertex anomalous dimension. 
This has led to the computation of the critical exponents to three orders in 
the large $N$ expansion, \cite{60,61,62,63,64,65}, in the underlying universal 
theory. In this instance, the four dimensional theory which is in the same 
universality class as the two dimensional Gross-Neveu model is the 
Gross-Neveu-Yukawa theory which has been discussed in this context in 
\cite{39,69}. Specifically the Lagrangian is
\begin{equation}
L_{\mbox{\footnotesize{GN}}}^{(4)} ~=~ i \bar{\psi}^i \partialslash \psi^i ~+~
\frac{1}{2} \partial_\mu \sigma \partial^\mu \sigma ~+~
\frac{1}{2} g_1 \sigma \bar{\psi}^i \psi^i ~+~ \frac{1}{24} g_2^2 \sigma^4 
\label{laggn4}
\end{equation}
where there are two couplings and the quadratic term in $\sigma$ becomes a
kinetic term since $\sigma$ has a canonical dimension of $1$. The process to 
generalize the Gross-Neveu thread in the same spirit as the scalar theories is 
now apparent. One solves the relation for the vertex dimension, (\ref{gnuniq}),
by allowing for an alternative fermion kinetic term leading to
\begin{equation}
\tilde{\alpha} ~=~ \mu ~-~ n ~+~ \half \eta ~~~,~~~ 
\tilde{\beta} ~=~ 2 n ~+~ 1 ~-~ \eta ~-~ \chi 
\label{qexp}
\end{equation}
where $n$~$=$~$0$ corresponds to (\ref{laggn22}). However, there is a new
feature here in that the fermion kinetic term is formally different depending
on whether $n$ is even or odd as a consequence of the necessary 
$\gamma$-algebra. When $n$ is odd one has a product of even numbers of 
$\partialslash$ which therefore translate into kinetic terms involving only 
$\Box$. In other words one ends up with the Klein-Gordon version of a fermionic 
field and its higher $\Box$ generalizations. We will ignore these type of 
solutions as one in effect will reproduce results and Lagrangians similar to 
the scalar theory threads. Therefore the first thread above $n$~$=$~$0$ to have
a kinetic term involving $\partialslash$ is $n$~$=$~$2$ with the Lagrangian 
\begin{equation}
L_{\mbox{\footnotesize{GN}}}^{(6)} ~=~ 
i \bar{\psi}^i \partialslash \Box \psi^i ~+~
\frac{1}{2} g_1 \sigma \bar{\psi}^i \psi^i  ~-~ \frac{1}{2} \sigma^2 
\label{laggn66}
\end{equation}
which has critical dimension $6$. A similar kinetic term has appeared in a 
different context in \cite{40}. There a fermionic model which was constructed 
on the assumption of having the fermion kinetic term of (\ref{laggn66}) but
requiring the interacting theory of one fermion to be renormalizable in four
dimensions. This led to an {\em eight}-point fermion self-interaction, 
\cite{40}, as well as $4$- and $6$-point interactions. It was noted in 
\cite{40} that the coupling of the $8$-point interaction was asymptotically
free. Here our premise is to retain a quartic fermion interaction for each 
thread as that underlying universal theory is accessible via the large $N$ 
expansion. It is not clear if the fermionic theory of \cite{40} is amenable to 
the large $N$ methods of Vasil'ev et al, \cite{21,22,23}. From (\ref{laggn66}) 
one can build higher dimensional Lagrangians for each thread value $n$. For 
instance, the continuation of the $n$~$=$~$2$ thread gives Lagrangians 
$L_{\mbox{\footnotesize{GN}}}^{(6,d)}$ where we will only consider $d$ to be
even. For an odd $d$ one cannot have a $\sigma$ kinetic term as there would 
have to be an odd number of derivatives and it is not possible to have a
Lorentz singlet. As the $\sigma$ field of this thread has canonical dimension 
$3$ the theories with $d$~$=$~$8$ and $10$ are the same as 
$L_{\mbox{\footnotesize{GN}}}^{(6)}$ with only a modified $\sigma$ kinetic 
term. Equally for $d$~$=$~$12$ to $16$ quartic interactions in $\sigma$ will be
present and involve derivative couplings. Then when $d$~$=$~$18$ a $6$-point 
$\sigma$ interaction will be necessary for renormalizability in addition to the 
independent quartic $\sigma$ interactions. To summarize we have  
\begin{eqnarray}
L_{\mbox{\footnotesize{GN}}}^{(6,8)} &=& 
i \bar{\psi}^i \partialslash \Box \psi^i ~+~ 
\frac{1}{2} \partial_\mu \sigma \partial^\mu \sigma ~+~
\frac{1}{2} g_1 \sigma \bar{\psi}^i \psi^i \nonumber \\
L_{\mbox{\footnotesize{GN}}}^{(6,10)} &=& 
i \bar{\psi}^i \partialslash \Box \psi^i ~+~ 
\frac{1}{2} \left( \Box \sigma \right)^2 ~+~
\frac{1}{2} g_1 \sigma \bar{\psi}^i \psi^i \nonumber \\
L_{\mbox{\footnotesize{GN}}}^{(6,12)} &=& 
i \bar{\psi}^i \partialslash \Box \psi^i ~+~ 
\frac{1}{2} \left( \Box \partial^\mu \sigma \right)^2 ~+~
\frac{1}{2} g_1 \sigma \bar{\psi}^i \psi^i ~+~
\frac{1}{24} g_2^2 \sigma^4 \nonumber \\
L_{\mbox{\footnotesize{GN}}}^{(6,14)} &=& 
i \bar{\psi}^i \partialslash \Box \psi^i ~+~ 
\frac{1}{2} \left( \Box^2 \sigma \right)^2 ~+~
\frac{1}{2} g_1 \sigma \bar{\psi}^i \psi^i ~+~
\frac{1}{24} g_2^2 \sigma^2 \partial^\mu \sigma \partial_\mu \sigma 
\nonumber \\
L_{\mbox{\footnotesize{GN}}}^{(6,16)} &=& 
i \bar{\psi}^i \partialslash \Box \psi^i ~+~ 
\frac{1}{2} \left( \Box^2 \partial^\mu \sigma \right)^2 ~+~
\frac{1}{2} g_1 \sigma \bar{\psi}^i \psi^i ~+~
\frac{1}{24} g_2^2 \sigma^2 \left( \Box \sigma \right)^2 \nonumber \\
&& +~ \frac{1}{24} g_3^2 \left( \partial^\mu \sigma \partial_\mu \sigma 
\right)^2 ~+~ 
\frac{1}{24} g_4^2 \sigma^2 \left( \partial^\mu \partial^\nu \sigma \right)^2 
\nonumber \\
L_{\mbox{\footnotesize{GN}}}^{(6,18)} &=& 
i \bar{\psi}^i \partialslash \Box \psi^i ~+~ 
\frac{1}{2} \left( \Box^3 \sigma \right)^2 ~+~
\frac{1}{2} g_1 \sigma \bar{\psi}^i \psi^i ~+~
\frac{1}{24} g_2^2 \sigma^3 \, \Box^3 \sigma ~+~ 
\frac{1}{24} g_3^2 \sigma^2 \, \Box \sigma \, \Box^2 \sigma \nonumber \\ 
&& +~ \frac{1}{24} g_4^2 \sigma^2 \, \partial_\mu \partial_\nu \sigma 
\, \Box \partial^\mu \partial^\nu \sigma ~+~ 
\frac{1}{24} g_5^2 \sigma^2 \, 
\left( \partial_\mu \partial_\nu \partial_\rho \sigma \right)^2 ~+~ 
\frac{1}{24} g_6^2 \sigma^2 \, \Box \partial_\mu \sigma \, 
\Box \partial^\mu \sigma \nonumber \\
&& +~ \frac{1}{24} g_7^2 \sigma \left( \Box \sigma \right)^3 ~+~ 
\frac{1}{24} g_8^2 \sigma \, \partial_\mu \partial_\nu \sigma 
\, \partial^\nu \partial^\rho \sigma \, \partial^\mu \partial_\rho \sigma ~+~ 
\frac{1}{24} g_9^2 \sigma \, \Box \sigma \, \partial_\mu \partial_\nu \sigma 
\, \partial^\mu \partial^\nu \sigma \nonumber \\
&& +~ \frac{1}{40320} g_{10}^4 \sigma^6 
\label{laggnd6dn}
\end{eqnarray}
where there are parallels in structure to the $O(N)$ scalar theories but a 
jump in the complexity when the critical dimension is $18$. 

This exercise can be repeated for the non-abelian Thirring model which 
like the Gross-Neveu model has a quartic fermion interaction and is 
renormalizable in two dimensions. The Lagrangian is  
\begin{equation}
L_{\mbox{\footnotesize{NATM}}}^{(2)} ~=~ 
i \bar{\psi}^i \partialslash \psi^i ~+~
\frac{1}{2} g_1^2 \left( \bar{\psi}^i \gamma^\mu T^a \psi^i \right)^2  
\label{lagnatm2}
\end{equation}
where $T^a$ are the generators of the underlying non-abelian symmetry. It is
accessible to large $\Nf$ computations where $\Nf$ is the number of fermion
fields and this parameter is chosen partly not to be confused with the 
parameter associated with the non-abelian group but mainly because of the
connection of (\ref{lagnatm2}) with QCD. As noted in \cite{66} the non-abelian 
Thirring model is in the same universality class as QCD at the Wilson-Fisher 
fixed point accessible via the large $\Nf$ expansion. So there is a thread of 
theories parallel to the $O(N)$ scalar and Gross-Neveu ones. The connection is 
more evident by the introduction of a spin-$1$ auxiliary field in the adjoint 
representation of the non-abelian symmetry group since
\begin{equation}
L_{\mbox{\footnotesize{NATM}}}^{(2)} ~=~ 
i \bar{\psi}^i \partialslash \psi^i ~+~
g_1 \bar{\psi}^i T^a \gamma^\mu \psi^i A^a_\mu ~-~ \frac{1}{2}
A^a_\mu A^{a\,\mu} ~.
\label{lagnatm22}
\end{equation}
In our discussions we will use the shorthand designation of gluon for this
spin-$1$ adjoint field even though this is usually understood to be the quanta
of the strong force only in four dimensions. From (\ref{lagnatm22}) the core 
quark-gluon vertex of QCD is apparent and it is this interaction which forms 
the basis of the universality class containing four dimensional QCD and was 
examined further in \cite{20}. Where (\ref{laggn22}) and (\ref{lagnatm22})
differ is that the presence of the $\gamma$-matrix in the vertex means that 
that vertex does not have any uniqueness condition in order to be able to 
develop its conformal integration. From the point of view of carrying out large
$\Nf$ computations this is not a limitation. Uniqueness can be exploited 
explicitly within the evaluation of contributing Feynman diagrams once the 
integrals are rewritten from tensor integrals to scalar ones. 

In parallel to (\ref{laggn66}) other threads of theories can be developed with 
higher derivative fermion kinetic terms. While the quark-gluon vertex of 
(\ref{lagnatm22}) is not unique the relation between the exponents of the 
fields 
\begin{equation}
2 \tilde{\alpha} ~+~ \tilde{\beta} ~=~ d ~+~ 1 ~-~ \chi ~.
\label{qalbetrel}
\end{equation}
is still valid from the scaling dimensions of the universal interaction where
$\chi$ here is the vertex anomalous dimension. Like the Gross-Neveu case the
solutions (\ref{qexp}) are still valid leading to a new thread of theories
based on the Lagrangian
\begin{equation}
L_{\mbox{\footnotesize{NATM}}}^{(6)} ~=~ 
i \bar{\psi}^i \partialslash \Box \psi^i ~+~
g_1 \bar{\psi}^i T^a \gamma^\mu \psi^i A^a_\mu ~-~ \frac{1}{2}
A^a_\mu A^{a\,\mu} 
\label{lagnatm66}
\end{equation}
where $A^a_\mu$ is to be regarded as an auxiliary field rather than a gauge
field. By this we mean that there is no restriction on the number of degrees of
freedom of the components $A^a_\mu$. In other words at this stage we are only 
interested in the structure of the Lagrangians of higher dimensional theories
in this $n$~$=$~$2$ thread. For instance, each interaction will carry its own
independent coupling constant which is not related to any others as would be
the case in a four dimensional gauge theory. In that case the gauge symmetry 
would place conditions on the coupling constants which would be preserved in 
the quantum theory via the Slavnov-Taylor identities. So, for instance, the
first few Lagrangians in the non-abelian Thirring model thread with $n$~$=$~$2$
are 
\begin{eqnarray}
L_{\mbox{\footnotesize{NATM}}}^{(6,8)} &=&
i \bar{\psi}^i \partialslash \Box \psi^i ~+~
g_1 \bar{\psi}^i T^a \gamma^\mu \psi^i A^a_\mu ~+~ 
\frac{\beta}{2} \partial_\mu A^a_\nu \, \partial^\mu A^{a\,\nu} ~+~ 
\frac{1}{2} (1-\beta) \partial^\mu A^a_\mu \, \partial_\nu A^{a\,\nu} 
\nonumber \\
L_{\mbox{\footnotesize{NATM}}}^{(6,10)} &=&
i \bar{\psi}^i \partialslash \Box \psi^i ~+~
g_1 \bar{\psi}^i T^a \gamma^\mu \psi^i A^a_\mu ~+~ 
\frac{\beta}{2} \partial_\mu \partial_\nu A^a_\sigma \, 
\partial^\mu \partial^\nu A^{a\,\sigma} ~+~ 
\frac{1}{2} (1-\beta) \partial^\mu \partial^\nu A^a_\nu \, 
\partial_\mu \partial_\sigma A^{a\,\sigma} \nonumber \\
&& +~ g_2 f^{abc} A^a_\mu A^b_\nu \partial^\mu A^{c\,\nu} ~+~ 
g_3 d^{abc} A^{a\,\mu} A^b_\mu \partial_\nu A^{c\,\nu} 
\end{eqnarray}
where $f^{abc}$ are the structure constants of the non-abelian Lie group and
$d^{abc}$ is the associated rank $3$ symmetric tensor. 

The parameter $\beta$ is not the gauge parameter of the linear covariant gauge 
fixing in a gauge theory as such. It should be regarded as an interpolating 
parameter. A parallel situation was noted in \cite{20} as being possible in the
quadratic part of the construction of the six dimensional Lagrangian which 
extends the NATM-QCD Wilson-Fisher fixed point equivalence to six dimensions. 
However in that situation the use of Bianchi identities for the field strength
and the gauge symmetry meant that the Lagrangian could be written in terms of 
one gauge independent $2$-leg gluonic operator. A second quadratic operator in 
the gluon was, however, necessary to effect the gauge fixing. In 
$L_{\mbox{\footnotesize{NATM}}}^{(6,d)}$ the parameter $\beta$ would become 
related to a gauge parameter if a higher symmetry was imposed on the
construction. To construct the next Lagrangian in the sequence, 
$L_{\mbox{\footnotesize{NATM}}}^{(6,12)}$, is a relatively straightforward
exercise in principal but requires an interplay between the tensors of the 
colour group and the structure of the interactions. For instance, 
$L_{\mbox{\footnotesize{NATM}}}^{(6,12)}$ will have quartic interactions in
addition to cubic ones in the gluon. The latter will however involve three
derivatives and involve both $f^{abc}$ and $d^{abc}$ tensors. Therefore the
number of independent interactions which are present in  
$L_{\mbox{\footnotesize{NATM}}}^{(6,12)}$ will be significantly larger 
especially since products of the two colour tensors will be possible. To gauge
the potential structure of such a Lagrangian, similar interactions arise in
four dimensional Yang-Mills theory when there is a nonlinear gauge fixing which
generalizes the 't Hooft-Veltman gauge of QED, \cite{70,71}. While this 
ultimately leads to a twelve dimensional Lagrangian with a significantly large 
number of interactions it would seem that the connection of the original 
non-abelian Thirring model with four dimensional gauge theory is not present in
the $n$~$=$~$2$ thread. This is apparent in the higher dimensional extensions 
of (\ref{lagnatm66}) where the square of the field strength does not appear to 
emerge even allowing for relations between coupling constants. However, we 
emphasise that our construction is based on the premises of requiring 
renormalizability with a core interaction which seeds the equivalence of 
theories at the Wilson-Fisher fixed point. There may be other fixed points
where such operators emerge naturally. 

\sect{Lower dimension completeness.}

While we have demonstrated that there is a large class of new renormalizable
quantum field theories which are the ultraviolet completeness of a higher
derivative $\phi^4$ theory one question which arises and which we can speculate
on is whether there is a set of theories {\em below} the critical dimension of 
the defining theory $L^{(D)}$ which we will call {\em lower dimension}
completeness. This is partly motivated by the fact that our $n$~$=$~$2$ and $3$
large $N$ critical exponents can in principle be expanded around dimensions
less than the critical dimension of $L^{(D)}$. Therefore, it is worth 
exploring the potential Lagrangians which could be present in the lower part of
the tower of theories for these new threads. Earlier we chose to begin with 
(\ref{lag4}) as the base Lagrangian. However it is well-known that that theory 
is in the same universality class as the nonlinear $\sigma$ model which is
\begin{equation}
L^{(4,2)} ~=~ \frac{1}{2} \left( \partial_\mu \phi^i \right)^2 ~+~ 
\frac{g_1}{2} \sigma \phi^i \phi^i ~-~ \frac{1}{2} \sigma  
\label{lag42} 
\end{equation}
in the notation we use here as it is renormalizable in two dimensions. While 
one could launch the tower of theories with this Lagrangian we chose not to do 
so as $L^{(4)}$ does not suffer from infrared issues. By this we mean that in 
two dimensions the basic $1/k^2$ bosonic propagator within a Feynman diagram is 
infrared divergent unlike in four dimensions. This is not unrelated to the fact
that in two dimensions $\phi^i$ is dimensionless. By contrast $L^{(4)}$ is
infrared finite. While (\ref{lag42}) is used for the large $N$ expansion, for 
perturbation theory the constraint is eliminated and the canonical form of the 
nonlinear $\sigma$ model is used which is
\begin{equation}
L^{(4,2)} ~=~ \frac{1}{2} g_{ab}(\pi) \partial^\mu \pi^a \partial_\mu \pi^b
\label{lag42p} 
\end{equation}
where $1$~$\leq$~$a$~$\leq$~$(N-1)$. For instance, in \cite{72} the
parametrization 
\begin{equation}
\phi^i ~=~ \left( \pi^a, \sqrt{\frac{1}{g_1} - \pi^b \pi^b } \, \right) 
\end{equation}
was used to establish the renormalization of (\ref{lag42p}). This version of 
the Lagrangian, (\ref{lag42p}), has a rich geometric structure as indicated by 
the presence of the metric of the underlying manifold. In this form there are 
an infinite number of interactions but one coupling constant with the theory
retaining its renormalizability. See, for instance, \cite{73}, for a review 
article. 

With this brief overview of the situation with the $n$~$=$~$1$ thread we can
now speculate on the potential for the lower dimensional completeness of the 
subsequent threads. For instance, the parallel to $L^{(4,2)}$ for $L^{(8)}$, 
(\ref{lag8}), would be 
\begin{equation}
L^{(8,4)} ~=~ \frac{1}{2} \left( \Box \phi^i \right)^2 ~+~ 
\frac{g_1}{2} \sigma \phi^i \phi^i ~-~ \frac{1}{2} \sigma ~. 
\end{equation}
Clearly this has similar infrared issues as $L^{(4,2)}$ since the $1/(k^2)^2$
propagator is not infrared safe in four dimensions. Again this is related to
the dimensionlessness of $\phi^i$ in this Lagrangian. The elimination of the
constraint does not lead to the same elegant geometric Lagrangian as 
$L^{(4,2)}$ due to the nature of the $2$-point term although there will be an
infinite number of interactions again. Despite this the theory should be 
renormalizable. This would need to be established prior to completing any
explicit computations. However in order to extract any renormalization group
functions one would have to add mass-like terms to regularize the infrared. One
such term cannot be $\phi^i \phi^i$ in $L^{(8,4)}$ since that disappears with
the elimination of the constraint. Instead terms such as 
$\frac{1}{2} m^2 \left( \partial_\mu \phi^i \right)^2$ could be used to
facilitate the infrared regularization. Such an analysis is beyond the scope
of the present article as our focus here is on the Wilson-Fisher fixed point
connections which can be accessed through the renormalization of {\it massless}
quantum field theories. Moreover, the verification of the lower dimension
completeness may not be as straightforward as the ultraviolet one. For 
instance, there is a clue in the large $N$ exponents $\eta^{(8)}$ and 
$\chi^{(8)}$ which we have computed here. If $L^{(8,4)}$ is the lower dimension
completeness of $L^{(8)}$ then the renormalization group functions must produce
exponents which agree with the $\epsilon$ expansion of the critical exponents 
of the eight dimensional large $N$ theory. Analysing these near four dimensions
indicates that there is a well-behaved $\epsilon$ expansion. This is not a 
minor point. For instance, in the extension of $L^{(8)}$ we have constructed 
theories in higher dimensions in steps of two. Equally there are parallels of 
the $n$~$=$~$2$ thread with the $n$~$=$~$1$ one. For the latter the large $N$ 
exponents of \cite{21,22} are divergent when one formally expands near 
{\em one} dimension which is one dimension below where $\sigma$ can appear 
linearly in the Lagrangian. The same situation arises for $n$~$=$~$2$ where the
corresponding large $N$ exponents diverge near three dimensions but have 
well-defined expansions for dimensions above three.

However, the dimensionalities of the fields and the parallel to $L^{(4)}$ 
directed us to $L^{(8,4)}$ which has skipped two dimensions. Examining the 
exponents $\eta^{(8)}$ and $\chi^{(8)}$ with respect to six dimensions produces
exponents which do not all begin with $O(\epsilon)$. This is similar to 
evaluating the exponents of \cite{21,22,23} in three dimensions. To understand 
this it is worth considering the potential form of a lower dimension
completeness to six dimensions. Based purely on the dimensionalities of the 
fields leads to
\begin{eqnarray}
L^{(8,6)} &=& \frac{1}{2} \left( \Box \phi^i \right)^2 ~+~
\frac{1}{2} \sigma \Box^{-1} \sigma ~+~
\frac{g_1}{2} \sigma \phi^i \phi^i ~+~ 
\frac{g_2^2}{4} \left( \phi^i \partial_\mu \phi^i \right)^2 \nonumber \\
&& +~ \frac{g_3^2}{2} \phi^i \phi^i \phi^j \Box \phi^j ~+~
\frac{g_4^4}{12} \left( \phi^i \phi^i \right)^3 
\label{lag86}
\end{eqnarray}
where the inverse box operator acts to the right and hence this Lagrangian 
contains a nonlocal quadratic term for $\sigma$. While such a construction is 
based on the same premises as those for the ultraviolet completion it is not 
clear whether this potential Lagrangian is the one which produces the critical 
exponents of $\eta^{(8)}$ and $\chi^{(8)}$ near six dimensions or not. If it 
does not or any modifications do not then there may be a doubt as to whether 
$L^{(8,4)}$ is part of the $n$~$=$~$2$ thread with a gap at six dimensions or 
whether there is a break below eight dimensions. In other words the lower
dimension completeness for $L^{(4,2)}$ is a special case. Similar remarks are 
applicable to the $n$~$=$~$3$ thread where 
\begin{equation}
L^{(12,6)} ~=~ \frac{1}{2} \left( \Box \partial_\mu \phi^i \right)^2 ~+~ 
\frac{g_1}{2} \sigma \phi^i \phi^i ~-~ \frac{1}{2} \sigma 
\end{equation}
would be the lower dimension completeness with the large $N$ exponents also
being divergent in $\epsilon$ near the odd dimensions five and below. Again in 
the critical dimension $\phi^i$ is dimensionless. So $L^{(12,6)}$ would also
require infrared regularization with mass-like terms in order to extract the 
renormalization group functions to compare with the large $N$ exponents. For 
the twelve dimensional universal theory these exponents have $\epsilon$ 
expansions which are $O(\epsilon)$ near six dimensions. However near ten 
dimensions $\chi^{(12)}_1$ has a pole in $\epsilon$ which does not have a 
parallel in the lower thread towers. This suggests some sort of pathology in 
the lower dimension completeness hypothesis in this and other intermediate 
theories which may be related to nonlocalities in the corresponding 
Lagrangians. Given this observation it may be the case that one has to consider
only local Lagrangians in the lower dimension completeness construction. As a 
side remark we note that like $n$~$=$~$1$ the next threads have finite 
exponents in odd dimensions. For instance,
\begin{eqnarray}
\left. \eta^{(8)} \right|_{d=7} &=& -~ \frac{256}{315\pi^2N} ~-~ 
\frac{1960214528}{31255875\pi^4N^2} ~+~ 
O \left( \frac{1}{N^3} \right) \nonumber \\
\left. \chi^{(8)} \right|_{d=7} &=& \frac{15872}{315\pi^2N} ~+~
O \left( \frac{1}{N^2} \right) 
\end{eqnarray}
and
\begin{eqnarray}
\left. \eta^{(12)} \right|_{d=11} &=&
\frac{131072}{405405\pi^2N} ~-~ 
\frac{569571965136797696}{7403290525756125\pi^4N^2} ~+~ 
O \left( \frac{1}{N^3} \right) \nonumber \\
\left. \chi^{(12)} \right|_{d=11} &=& \frac{65536}{567\pi^2N} ~+~
O \left( \frac{1}{N^2} \right) ~. 
\end{eqnarray}

Nonlocalities, however, can be accommodated within a local quantum field theory
context if the nonlocality can be localized. This is the case for the Gribov 
operator in QCD, \cite{74}, where the operator was localized by introducing 
localizing ghost fields, \cite{75,76,77,78,79,80}. The resulting 
Gribov-Zwanziger Lagrangian was shown to be renormalizable and amenable to 
multiloop renormalization. Again there are parallels here since $L^{(8,6)}$ can
be localized in the same vein to give
\begin{eqnarray}
L^{(8,6)} &=& \frac{1}{2} \left( \Box \phi^i \right)^2 ~+~ 
\frac{1}{2} \rho \Box \rho ~+~ \rho \sigma \nonumber \\
&& +~ \frac{g_1}{2} \sigma \phi^i \phi^i ~+~ 
\frac{g_2^2}{4} \left( \phi^i \partial_\mu \phi^i \right)^2 ~+~ 
\frac{g_3^2}{2} \phi^i \phi^i \phi^j \Box \phi^j ~+~
\frac{g_4^4}{12} \left( \phi^i \phi^i \right)^3 \nonumber \\
&& +~ \frac{g_5}{6} \rho^3 ~+~ \frac{g_6^2}{4} \rho^2 \phi^i \phi^i ~+~
\frac{g_7}{2} \rho \partial^\mu \phi^i \partial_\mu \phi^i ~+~
\frac{g_8}{2} \rho \phi^i \Box \phi^i ~. 
\label{lag86loc}
\end{eqnarray}
We have followed the localization prescription used in the construction of the 
Gribov-Zwanziger Lagrangian, \cite{75,76,77,78,79,80}. In this an auxiliary 
field, which in our case is $\rho$, is initially introduced in such a way that 
when its equation of motion is derived in (\ref{lag86loc}) and included in the 
kinetic part of the Lagrangian then the nonlocal $\sigma$ $2$-point term of 
(\ref{lag86}) is produced. More specifically the $\rho$ equation of motion 
produces
\begin{equation}
\rho ~=~ \Box^{-1} \sigma ~.
\end{equation}
In (\ref{lag86loc}) the original interactions are retained. However, with the 
introduction of a new field there is the potential to have new interactions. 
These are represented by additional terms in (\ref{lag86loc}) with couplings 
$g_5$ to $g_8$ and there are no such corresponding terms present in the 
original Lagrangian of (\ref{lag86}) as it stands. These interactions arise 
because $\rho$ has dimension $1$ and therefore on power counting grounds they 
would be required to ensure renormalizability. In other words if we ignore for 
the moment the notion of lower dimension completeness of $L^{(8)}$ and one were 
given scalar fields $\phi^i$, $\sigma$ and $\rho$ of respective canonical 
dimensions $1$, $4$ and $2$ then (\ref{lag86loc}) would be the most general 
renormalizable quantum field theory one could construct in the critical 
dimension six. The $\sigma$-$\rho$ $2$-point term is present on dimensional 
grounds. While such a $2$-point term may appear to be unnatural in a Lagrangian
a term with a similar structure is present in the localized renormalizable 
Gribov-Zwanziger Lagrangian, \cite{75,76,77,78,79,80}. Specifically there is a 
$2$-point term involving the spin-$1$ adjoint and the bosonic localizing ghost 
fields. While it leads to a matrix of propagators this is not a hindrance to 
performing perturbative computations. In other words one can regard 
(\ref{lag86loc}) as a local renormalizable Lagrangian with which to perform 
computations. Whether its renormalization group functions and resulting 
critical exponents then have any connection to the underlying universal theory 
of the $L^{(8)}$ thread is not clear and beyond the scope of the present 
article. 

Given this excursion to discuss the local Lagrangian (\ref{lag86loc}) it is 
worth reviewing (\ref{lag86}) in light of what we have constructed. Our 
Lagrangian (\ref{lag86}) was written down purely on the grounds of the 
dimensionality of the fields and the ethos of one common force-matter 
interaction corresponding to the interaction with coupling $g_1$. The other 
underlying assumption was one of locality but that has been loosened a little 
by the presence of the nonlocal kinetic term. The absence of couplings 
corresponding to $g_5$ to $g_8$ in (\ref{lag86loc}) may appear to be 
inconsistent. However, in light of (\ref{lag86loc}) and completely dropping the
locality assumption would lead to 
\begin{eqnarray}
L^{(8,6)}_{\mbox{\footnotesize{nl}}} &=& 
\frac{1}{2} \left( \Box \phi^i \right)^2 ~+~ 
\frac{1}{2} \sigma \Box^{-1} \sigma ~+~ \frac{g_1}{2} \sigma \phi^i \phi^i ~+~ 
\frac{g_2^2}{4} \left( \phi^i \partial_\mu \phi^i \right)^2 ~+~ 
\frac{g_3^2}{2} \phi^i \phi^i \phi^j \Box \phi^j \nonumber \\
&& +~ \frac{g_4^4}{12} \left( \phi^i \phi^i \right)^3 \nonumber ~+~ 
\frac{\tilde{g}_5}{6} \left( \Box^{-1} \sigma \right)^3 ~+~  
\frac{\tilde{g}_6^2}{4} \left( \Box^{-1} \sigma \right)^2 \phi^i \phi^i ~+~
\frac{\tilde{g}_7}{2} \left( \Box^{-1} \sigma \right) \partial^\mu \phi^i 
\partial_\mu \phi^i \nonumber \\
&& +~ \frac{\tilde{g}_8}{2} \left( \Box^{-1} \sigma \right) \phi^i \Box 
\phi^i 
\label{lag86nl}
\end{eqnarray}
on dimensional grounds. We have introduced parallel couplings $\tilde{g}_5$ to
$\tilde{g}_8$ as they are not necessarily equivalent to those of 
(\ref{lag86loc}). Again it is not clear if this is the lower dimension complete
Lagrangian for the $L^{(8)}$ thread in six dimensions. With the presence of
nonlocal interactions now it is not clear how (\ref{lag86nl}) can be localized.
Like the nonlocal $2$-point term of (\ref{lag86}) this more nonlocal
Lagrangian has parallels in four dimensional gauge theories. For instance, the
gauge invariant nonlocal mass operator
\begin{equation}
{\cal O} ~ \equiv ~ \frac{1}{2} \stackrel{\mbox{\begin{small}min\end{small}}}
{\mbox{\begin{tiny}$\{U\}$\end{tiny}}} \int d^4x \, \left( A^{a \, U}_\mu
\right)^2 
\label{a2nl}
\end{equation}
has nonlocal cubic terms and higher in its coupling constant expansion. See,
for example, the review article \cite{81} for more background. Here $A^U_\mu$ 
is a gauge invariant gauge field by construction and
\begin{equation}
A^U_\mu ~=~ U A_\mu U^\dagger ~-~ \frac{i}{g} \left( \partial_\mu U \right)
U^\dagger ~.
\end{equation}
Moreover, these can be ordered into a series of gauge invariant operators,
\cite{82,83}. For instance,
\begin{eqnarray}
{\cal O} &=& -~ \frac{1}{2} G^{a \, \mu\nu} \frac{1}{D^2} G^{a \, \mu\nu}
\nonumber \\
&& +~ g f^{abc} \left( \frac{1}{D^2} G^{a \, \mu\nu} \right)
\left( \frac{1}{D^2} D^\sigma G^b_{\sigma\mu} \right)
\left( \frac{1}{D^2} D^\rho G^c_{\rho\nu} \right) \nonumber \\
&& -~ g f^{abc} \left( \frac{1}{D^2} G^{a \, \mu\nu} \right)
\left( \frac{1}{D^2} D^\sigma G^b_{\sigma\rho} \right)
\left( \frac{1}{D^2} D^\rho G^c_{\mu\nu} \right) ~+~ O(g^2) 
\end{eqnarray}
where $G^a_{\mu\nu}$ is the non-abelian field strength and $D_\mu$ is the
covariant derivative. The expansion has a structure which is similar to the
nonlocal interactions in (\ref{lag86nl}). The sticking point of treating the
gauge invariant mass operator is that the localization of a nonlocal $3$-point 
interaction has not been established yet for (\ref{a2nl}) as well as the 
infinite number of nonlocal operators in the series. So to ascertain whether 
(\ref{lag86nl}) is in fact the lower dimension completeness of $L^{(8)}$ is not 
immediately possible. In discussing the construction of (\ref{lag86}) and its 
potential completeness another assumption has been implicitly dropped in 
(\ref{lag86nl}). That is that there is only one force-matter interaction which 
is the cubic one. In (\ref{lag86nl}) as well as in (\ref{lag86loc}) by 
extension there are now additional cubic as well as quartic force-matter 
interactions. The upshot of naturally examining what we have termed the lower 
dimension completeness of the $L^{(8)}$ thread is to open up a more complex set 
of potential Lagrangians which would need to be analysed and is beyond the 
scope of the present article. 

\sect{Discussion.}

One of the main features of our investigation of the $O(N)$ scalar field 
theories is the observation that the universal theory based on a $\phi^4$
interaction has an infinite number of universality classes. The core 
force-matter interaction, $\sigma \phi^i \phi^i$, defines the linear relation 
between the dimensions of the separate fields. In \cite{21,22,23} a specific
solution was examined at length which we have denoted by the $n$~$=$~$1$ thread
here. In some sense one ordinarily regards the kinetic terms as the canonical 
starting point for constructing a Lagrangian rather than the interaction. In 
the way we have considered the Lagrangian construction from a critical point 
perspective the interaction by contrast informs the kinetic term. The $n$ 
referred to relates to or classifies the powers of derivatives in the kinetic 
term. For integer values of $n$~$>$~$1$ higher derivative kinetic terms emerge.
While this increases the critical dimension which a Lagrangian is 
renormalizable in it opens up a host of new Lagrangians which can be studied 
within the developing $d$-dimensional conformal field theory formalism. While 
free field higher derivative kinetic terms have been investigated in 
\cite{29,30,31}, for instance, there is now an opportunity to study interacting
cases. This can be used as a new laboratory to study connections with the 
AdS/CFT ideas as well as a starting point to classify and more importantly 
connect scalar quantum field theories. Moreover, given that our initial 
motivation was in $O(N)$ scalar theories we have shown that the higher threads 
of $n$ are accessible via the large $N$ expansion technique developed in 
\cite{21,22,23}. In addition we have constructed the ultraviolet completions of
several of the theories in each thread and shown by perturbative analysis that 
they do indeed lie in the same universality class. These are nontrivial checks 
and required the use of various connecting techniques such as the Tarasov 
method for relating $d$-dimensional Feynman integrals with similar integrals in
$(d+2)$-dimensions. We also had to compute new large $N$ $d$-dimensional 
critical exponents for the $n$~$=$~$2$ and $3$ threads as these have to be in 
total agreement with the perturbative renormalization group functions of the 
fixed dimension Lagrangians lying in the tower of each thread. The next stage 
in this will be the computation of other large $N$ critical exponents such as 
$\nu$ as well as $\eta$ at $O(1/N^3)$. The latter in the $n$~$=$~$1$ thread 
used the early conformal bootstrap method of \cite{23}. Equally we have 
concentrated on the $n$~$=$~$2$ and $3$ threads but there is no reason why the 
analysis we have given here cannot be extended aside from the practical 
computation limitation which we encountered. 

With the core $\phi^4$ Lagrangians in the $n$~$>$~$1$ threads having a critical
dimension greater than $4$ there is a new potential feature which is what we 
termed lower dimension completeness. While this is more speculative as to
whether there are connecting Lagrangians in the same universality class the 
complicating feature appears to be the presence of nonlocalities. At a critical
point this would not be as major an issue as trying to construct a viable 
nonlocal Lagrangian away from criticality. There are examples, such as that of 
Gribov, \cite{74}, which can be renormalized after the localization process 
introduced by Zwanziger, \cite{75,76,77,78,79,80}. In principle this provides a
potential route to study lower dimension complete Lagrangians and we have 
suggested a toy model in our $L^{(8,6)}$ hypothesis as a place to begin. 
Understanding nonlocalities in a Lagrangian context may inform models of colour
confinement in Yang-Mills theories for which the Gribov construction has 
already been widely studied. However, this will require going beyond the scalar
theories considered in the main part of the article. As the key connecting tool
across the dimensions appears to be the large $N$ expansion we briefly 
discussed the development of the scalar theory ideas in fermionic models such 
as the $O(N)$ Gross-Neveu model and the non-abelian Thirring model. In each 
there is a parallel defining relation between the dimensions of the matter and 
force fields which admit higher derivative solutions. The extension of the 
ideas to these fermionic theories has yet to be analysed in the same depth 
perturbatively or in the large $N$ construction which we hope to examine in 
future work. 

\vspace{1cm}
\noindent
{\bf Acknowledgements.} One of the authors (JAG) thanks Prof. D. Kreimer for 
valuable discussions as well as the Kolleg Mathematik Physik Berlin for
financial support. The Mathematical Physics Group at Humboldt University, 
Berlin, where part of the work was carried out, is also thanked for its 
hospitality. The work was carried out with the support of the STFC through the
Consolidated Grant ST/L000431/1 and a studentship (RMS).

\end{document}